\documentclass{aa}

\usepackage{graphics}
\usepackage{epsfig,psfig}

\newcommand{\nodata}{\multicolumn{2}{c}{--}}
\newcommand{\rb}{\raisebox{-1.0mm}}

\begin{document}

\title{Study of the properties and spectral energy distributions of the
      Herbig AeBe stars HD 34282 and HD 141569 \thanks{Based on observations
    made with the CST, NOT, INT and WHT telescopes of the Canary Islands
    observatories under the auspices of its International Time Programme.
    Also based on observations made with the 2.2m telescope at Calar Alto
    Observatory (Almer\'{\i}a).}}

\author{B. Mer\'\i n\inst{1}, B. Montesinos\inst{1,2}, C. Eiroa\inst{3}, E.
  Solano\inst{1}, A. Mora\inst{3}, P. D'Alessio\inst{4}, N. Calvet\inst{5},
  R.D. Oudmaijer\inst{6}, D. de Winter\inst{7}, J.K. Davies\inst{8}, A.W.
  Harris\inst{9}, A. Cameron\inst{10}, H.J. Deeg\inst{11}, R.
  Ferlet\inst{12}, F. Garz\'on\inst{11}, C.A. Grady\inst{13}, K.
  Horne\inst{10}, L.F. Miranda\inst{2}, J. Palacios\inst{3}, A.
  Penny\inst{14}, A. Quirrenbach\inst{15}, H. Rauer\inst{9}, J.
  Schneider\inst{16}, P.R. Wesselius\inst{17}}

\offprints{Bruno Mer\'{\i}n, e-mail address: bruno@laeff.esa.es}

\institute{Laboratorio de Astrof\'{\i}sica Espacial y F\'{\i}sica Fundamental (LAEFF), 
Apartado 50727, 28080 Madrid, Spain
\and
Instituto de Astrof\'\i sica de Andaluc\'\i a-CSIC,
Apartado 3004, 18080 Granada, Spain
\and
Departamento de F\'\i sica Te\'orica, M\'odulo C-XI, Facultad de 
Ciencias, Universidad Aut\'onoma de Madrid, 28049 Cantoblanco, Madrid, Spain 
\and
Centro de Radioastronom\'{\i}a y Astrof\'{\i}sica (UNAM), Apartado Postal 3-72 (Xangari)
58089, Morelia, M\'exico.
\and
Harvard-Smithsonian Center for Astrophysics, 60 Garden Street, Cambridge, MA 02138, USA
\and
Department of Physics and Astronomy, University of Leeds,
Leeds LS2 9JT, UK
\and
TNO/TPD-Space Instrumentation, Stieltjesweg 1, PO Box 155,
2600 AD Delft, The Netherlands
\and
Astronomy Technology Centre, Royal Observatory, Blackford Hill, Edinburgh, EH9 3HJ, UK
\and
DLR Department of Planetary Exploration, Rutherfordstrasse 2,
12489 Berlin, Germany
\and
Department of Physics and Astronomy, University of St. Andrews,
North Haugh, St. Andrews KY16 9SS, Scotland, UK
\and
Instituto de Astrof\'\i sica de Canarias, c/Via L\'actea s/n, 38200 
La Laguna, Tenerife, Spain
\and
CNRS, Institute d'Astrophysique de Paris, 98bis Bd. Arago,
75014 Paris, France
\and
NOAO/STIS, Goddard Space Flight Center, Code 681, NASA/GSFC,
Greenbelt, MD 20771, USA
\and
Rutherford Appleton Laboratory, Chilton, Didcot, Oxfordshire OX11 0QX, UK
\and
Sterrewacht Leiden, PO Box 9513, 2300 RA Leiden, The Netherlands
\and
Observatoire de Paris, place Jules Janssen, 92195 Meudon, France
\and
SRON, Universiteitscomplex ``Zernike", Landleven 12,
P.O. Box 800, 9700 AV Groningen, The Netherlands
}

\date{Received --, accepted -- }

\abstract{We present a study of the stellar parameters, distances and
  spectral energy distributions (SEDs) of \object{HD 34282} and
  \object{HD 141569}, two pre-main sequence Herbig AeBe stars.  Both
  objects have been reported to show `anomalous positions' in the HR
  diagram in the sense that they appear below the main sequence.  A
  significant result of this work is that both stars are
  metal-deficient.  The {\it Hipparcos} distance of HD 34282 is very
  uncertain and the current study places the star at the expected
  evolutionary position in the HR diagram, i.e. as a PMS star. The
  distance for HD 141569 found in this work matches the {\it
  Hipparcos} distance, and the problem of its anomalous position is
  solved as a result of the low metallicity of the object: using the
  right metallicity tracks, the star is in the PMS region.  The SEDs
  are constructed using data covering ultraviolet to millimetre
  wavelengths. Physical, non-parametric models, have been applied in
  order to extract some properties of the disks surrounding the
  stars. The disk around HD 34282 is accreting actively, it is massive
  and presents large grains in the mid-plane and small grains in the
  surface. HD 141569 has a very low mass disk, which is in an
  intermediate stage towards a debris-type disk.  }

\titlerunning{A study of the properties and SEDs of HD 34282 and HD 141569}
\authorrunning{Mer\'{\i}n et al.}

\maketitle

\section{Introduction}
\label{THEINTRO}

The study of protoplanetary disks is currently undergoing an exciting
stage partly propelled by the discovery of extrasolar planetary
systems following the detection of 51 Peg B by Mayor \& Queloz (1995).
The possible discovery of telluric planets in the near future, in
addition to the jovian-like planets already detected, will pose
interesting questions on the formation of extrasolar planetary
systems. Knowledge of the properties of protoplanetary disks and how
they evolve to debris disks around main-sequence (MS) stars is one of
the tools required for modelling such a process.

Observed spectral energy distributions (SEDs) of pre-main sequence
(PMS) stars are widely used to study the properties of protoplanetary
disks and to classify T Tauri and Herbig AeBe stars (HAeBe hereafter)
into an evolutionary scheme (Adams et al. 1987; Hillenbrand et al.
1992). Although the interpretation of a given SED is, to some extent,
model dependent, the theoretical modelling of SEDs constitutes an
invaluable tool for understanding the structure and properties of
protoplanetary disks, e.g.  Chiang \& Goldreich (1997, 1999),
D'Alessio et al. (1998, 1999, 2001), Dullemond et al.  (2001).  The
EXPORT consortium (Eiroa et al. 2000) observed a large sample of PMS
and Vega-type stars during the 1998 International Time Programme of
the Canary Islands' Observatories. One of the driving goals of this
effort was the study of the evolution and properties of protoplanetary
disks by analysing the SEDs of the young stars, taking advantage of
the fact that the EXPORT optical and near-IR photometry were obtained
simultaneously.  This observational approach is appropriate since T
Tauri and HAeBe stars vary markedly in these spectral regimes and a
significant part of the total luminosity of the object is radiated by
the PMS stellar photosphere at these wavelengths.  Among the stars in
the EXPORT sample with measured {\it Hipparcos} parallaxes, the HAeBe stars
HD 34282 and HD 141569 are the only ones whose positions fall below
the zero-age main sequence in a $\log L - \log T_{\rm eff}$ HR diagram
or equivalent (e.g.  van den Ancker et al. 1998, Weinberger et al.
2000).  This result is difficult to reconcile with some observational
results and with theoretical PMS evolutionary models (e.g.  Yi et al.
2001).

In this paper we present an analysis of the stellar properties of HD 34282
and HD 141569 and their circumstellar disks, based on EXPORT data
complemented with new spectroscopic observations and data from the
literature.  The structure of the paper is as follows. In Section
\ref{THESTARS} we briefly review what is currently known about the stars.
Sections \ref{THEOBS} and \ref{THERESULTS} present the observations used in
this work and their results. In Section \ref{THEPARAMS} we do a
comprehensive study of the stellar parameters and distances to the stars.
In Section \ref{THESEDS} the disk models used to reproduce the observed SEDs
are presented. In Section \ref{THECONC} we summarize the results of the
work.

\section{The stars}
\label{THESTARS}

{\it HD 34282:} It is a HAeBe star with spectral type estimates in the
range A0~V - A3~V and a rotational velocity of $v \sin i\!=\!129\pm 8$
km/s (Mora et al. 2001).  It is variable with published $m_V$ values
between 9.8 and 10.11 mag (Sylvester et al.  1996; de Winter et al.
2001).  Malfait et al.  (1998) report a visual variability amplitude
of $\sim$ 2.5 mag, while an Algol-type minimum of $\Delta m\sim$0.8
mag is observed in the {\it Hipparcos} light curve, indicating that HD 34282
is most likely an UXOR-type object. Its SED shows a strong IR excess
which is already noticeable at near-IR wavelengths (Sylvester et al.
1996; Malfait et al.  1998).  The {\it Hipparcos} parallax is
$\pi\!=\!6.10\pm 1.63$ mas, equivalent to a distance of
$164^{+60}_{-30}$ pc. Based on this distance, van den Ancker et al.
(1998) estimate a luminosity of 4.8 $L_\odot$, a value which locates
HD 34282 in an anomalous position in the HR diagram, well below the
expected luminosity of a star of its spectral type.  Note that the
error in the parallax, namely $\sigma\!=\!1.63$ mas, is slighly larger
than usual for {\it Hipparcos} objects, most probably due to the
faintness of the star, affecting the estimate of the distance. A large
keplerian disk around HD 34282 has been inferred from interferometric
$^{12}$CO $J\!=\!2\!\rightarrow\!1$ line observations (Pi\'etu at al.
2003).  Those authors estimate a distance $d\!=\!400^{+170}_{-100}$ pc
based on dynamical considerations, which yields
$L_*/L_\odot\!=\!29^{+30}_{-13}$ and
$M_*/M_\odot\!=\!2.1^{+0.4}_{-0.2}$ for the star, much more consistent
with the expected stellar properties of HD 34282.

{\it HD 141569:} This star is considered to be a transitional object
between the PMS HAeBe stars and the more evolved Vega-type systems.
Optical and near-IR images reveal a circumstellar disk with a complex
annular structure extending up to a radial distance of about 600 AU
(Weinberger et al.  1999; Augereau et al. 1999; Mouillet et al. 2001;
Clampin et al. 2003).  Mid-IR images trace the inner portion of the
disk, up to about 100 AU (Fisher et al. 2000; Marsh et al.  2002). The
disk in HD 141569 shares many of the properties of dusty debris disks
associated with young MS stars, such as $\beta$ Pic or \object{HR
4796}. On the other hand, the gas content of the disk is significantly
higher than that of MS debris disks, as evidenced by the CO
measurements at millimetre and near-IR wavelengths (Zuckerman et al.
1995; Brittain \& Rettig 2002; Boccaletti et al. 2003).

\begin{table*}
\caption[]{Telescopes, instruments, setups and observing modes.}
\begin{tabular}{lllll}
Telescope   & Instrument              & Range/Bands       & $\lambda/\Delta\lambda$  & Observing mode\\\hline
WHT (4.2m)  & UES                     & 3800--5900 \AA    & $ \sim\!49000$           & Echelle spectroscopy \\
INT (2.5m)  & IDS                     & 5800--6800 \AA    & $ \sim\!5000$            & Mid resolution spectroscopy\\
NOT (2.5m)  & Turpol (photopol.)      & {\it UBVRI}       &                          & Photopolarimetry\\
CST (1.5m)  & Photometer+near-IR camera &     {\it JHK}   &                          & Photometry    \\
CAHA 2.2m   & CAFOS (spectros.)       & 3700--6440 \AA    & $ \sim\!2400$            & Mid resolution spectroscopy\\\hline
\end{tabular} 

\vspace*{0.4cm}
\begin{tabular}{lll} 
\multicolumn{2}{l}{Observing dates for the spectroscopic observations$^\dag$} & \\
                 &  HD 34282                         & HD 141569             \\\hline
WHT/UES          &  28, 29 January 99                & 16, 17 May 98 (2+2)   \\  
                 &                                   & 28, 29, 30 July 98     \\
                 &                                   & 28, 29, 30, 31 January 99 \\ \hline
INT/IDS          &  24, 25, 26, 27, 28 October 98        & 14, 15, 16, 17 May 98 \\
                 &  29, 30, 31 January 99                & 29, 30, 31 July 98     \\     
                 &                                   & 29, 30, 31 January 99     \\ \hline
CAHA 2.2/CAFOS   &  3 February 03                         & 6 March 03              \\ \hline
\end{tabular}

$\dag$ The observing logs for the photometric observations are given by Eiroa et al.\\
(2001) and Oudmaijer et al. (2001).
\label{SETUPS}
\end{table*}

Concerning the star itself, HD 141569 has spectral type estimates between
B9 and A0V and a very high rotational velocity, $v \sin i\!=\!258\pm 17$
km/s (Andrillat et al. 1990; Mora et al. 2001 and references therein). The
star has no photometric variability (Alvarez \& Schuster 1981). Its
spectrum exhibits double-peaked H$\alpha$ emission superimposed on a broad
and strong absorption line (Andrillat et al.  1990; Dunkin et al. 1997) as
well as double-peaked emission of O {\sc i} at 7772 and 8446 \AA{}
(Andrillat et al.  1990); an outflow of material is suggested by the Na
{\sc i} D lines (Dunkin et al. 1997). Van den Ancker et al.  (1998)
estimate an age $>10^7$ years, while Weinberger et al.  (2000) give an age
of 5 Myr, on the basis of the stellar properties and age estimates of two
nearby stars, which are most likely its companions and have T Tauri
characteristics.  Considering the {\it Hipparcos} distance, 99 pc, and the
observed optical brightness and colours (e.g. de Winter et al. 2001), we
deduce an absolute magnitude of M$_V$ = 1.8 mag, assuming an A0V spectral
type and a typical interstellar extinction law (Rieke \& Lebofsky 1985).
This makes the star underluminous, a property seemingly shared by the
debris disk systems $\beta$ Pic, HR 4796 and some other A-type young MS
stars (Jura et al.  1998, Lowrance et al. 2000). We note, however, that
HD 141569 has the largest IR excess among all these objects, it is the only
object with emission lines in its optical spectrum, there is evidence of
outflowing gas, its disk has a remarkable gas content, and it likely has
the smaller evolutionary age. In addition, from a PMS evolution theory
point of view, a 5 Myr old A0 star ($T_{\rm eff}$ = 10000 K, see below)
with solar metallicity would have an absolute magnitude M$_V$ = 1.0--1.1
mag (e.g. Yi et al. 2001), which is significantly brighter than the
magnitude deduced from the {\it Hipparcos} parallax.

\section{Observations}
\label{THEOBS}

High resolution echelle (wavelength range 3800--5900 \AA) and intermediate
resolution (wavelength range 5800--6700 \AA) spectroscopy, optical
photo-polarimetry and near-infrared photometry of HD 34282 and HD 141569
were obtained during the four EXPORT observational campaigns in May, July
and October 1998 and January 1999 carried out at the Canary Islands
observatories (Eiroa et al. 2000).  Details of the
observations, instrumental setups and reduction procedures for each
particular observing mode are given in Eiroa et al. (2001),
Mora et al. (2001) and Oudmaijer et al.  (2001), so
we do not repeat them here, though Table \ref{SETUPS} provides an overview
of the telescopes, instruments, setups and observing modes used during the
EXPORT campaigns, plus the additional observations carried out in 2003 (see
below). The dates for the spectroscopic observations are also specified.

The spectra of HD 34282 and HD 141569 are presented and analysed here
for the first time, while the photometric data, though already
published as such in the previous references, are collected here
together and also discussed for the first time. We note that the data
are simultaneous or almost simultaneous, allowing a coherent study to
be made of the photometric and spectroscopic variations, and the
construction of the corresponding SEDs in the interval {\it UBVRIJHK}
with contemporaneous data.

In addition to the EXPORT observations, intermediate resolution
spectra in the range 3700--6200 \AA{}, with very high signal-to-noise
ratio, have been taken for both stars with CAFOS (Calar Alto Faint
Object Spectrograph) on the 2.2m telescope at Calar Alto observatory
(Almer\'{\i}a, Spain).  The motivation was to obtain high quality
Balmer line profiles, in particular H$\beta$, H$\gamma$ and H$\delta$,
in order to estimate the stellar gravities (see Section
\ref{THEPARAMS}). The echelle spectra were inadequate for analysis of
the Balmer line profiles: these are so wide that their wings extend up
to three echelle spectral orders -- the correction of the response
function of the spectrograph and the splicing of adjacent orders are
not accurate enough to consider the reconstructed profiles as
reliable.  CAFOS was equipped with a CCD SITe detector of
2048$\times$2048 pixels (pixel size 24 $\mu$m) and the grism Blue-100,
centered at 4238 \AA, giving a reciprocal linear dispersion of 88
\AA/mm (2 \AA/pixel).  Three spectra, with exposure times 120, 1200
and 1200 s, were taken for HD 34282 on 5 February 2003, and two with
exposure times 10 and 100 s were obtained for HD 141569 on 6 March
2003. The usual bias, dark and dome flat-field frames were taken in
each case. Standard procedures were used to process the data. The
nominal wavelengths of the Balmer lines were used to self-calibrate
the spectra in wavelength.

\section{Results}
\label{THERESULTS}

\subsection{Spectroscopy}

\begin{figure}
\epsfig{file=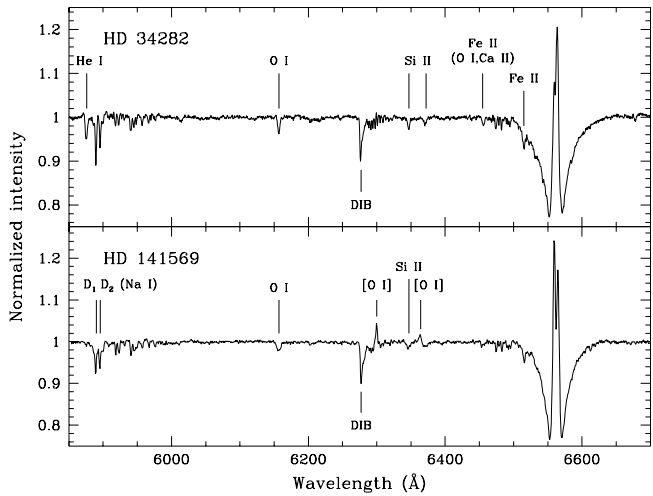,height=7.0cm}
\caption[]{The normalized average IDS/INT spectra of HD 34282 and HD 141569. 
Some prominent lines are identified. DIB stands for ``diffuse interstellar 
band''. The spectrum of HD 34282 is the average of the January 1999 run while
the spectrum of HD 141569 is the average of the May 1998 run.}
\label{INT}
\end{figure}

\begin{figure}
\epsfig{file=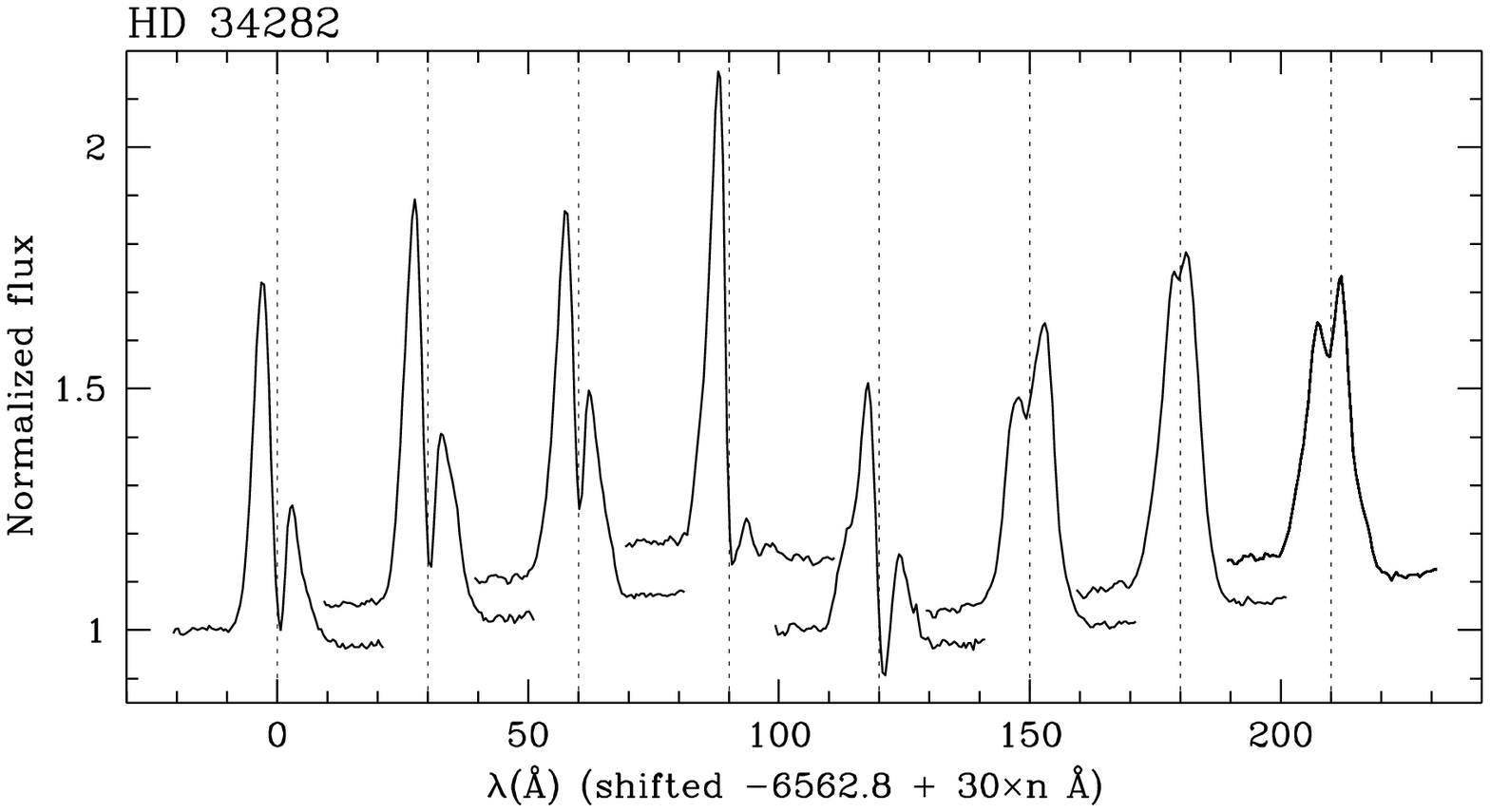,height=5.0cm}
\epsfig{file=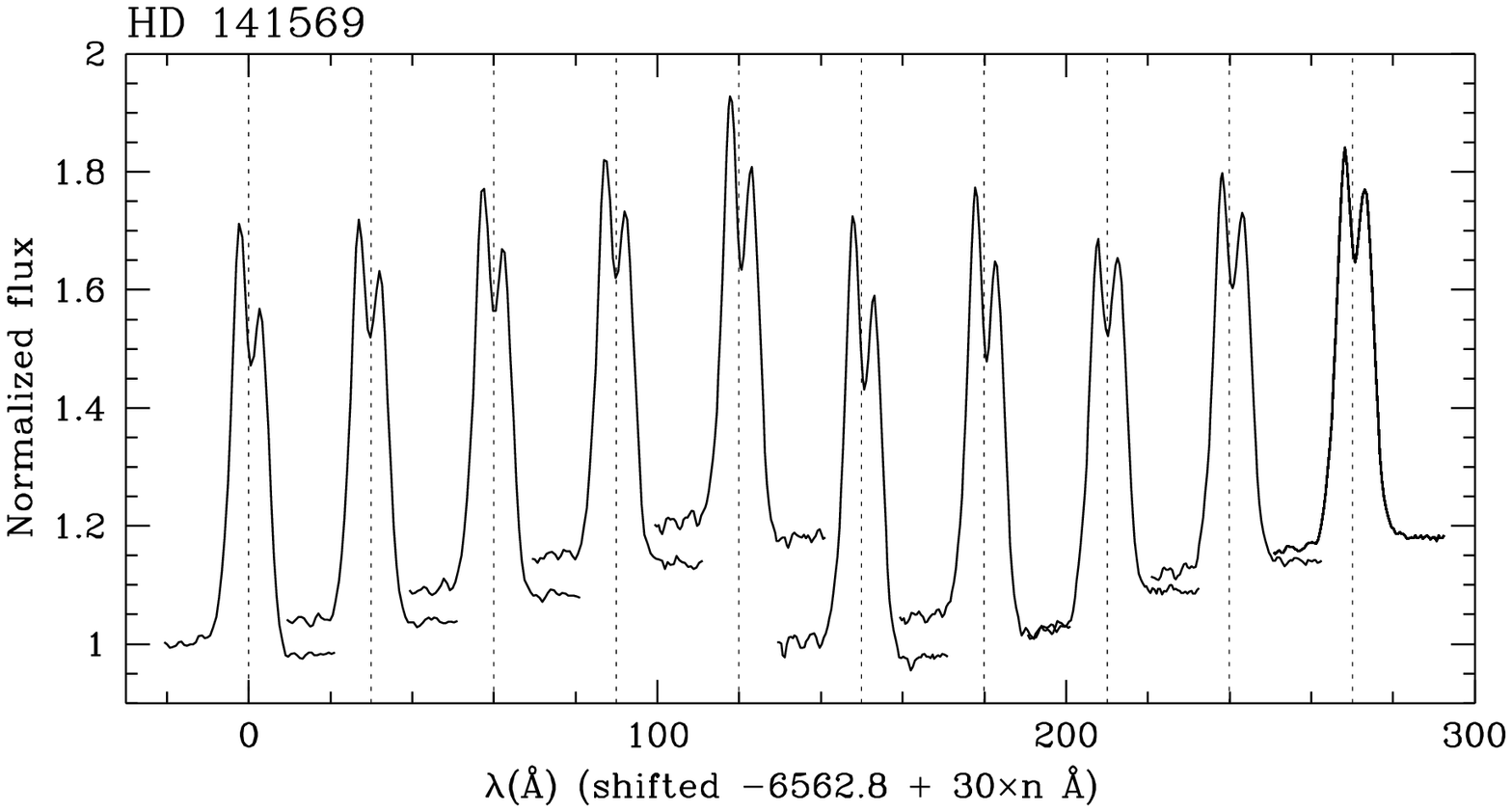,height=5.0cm}
\caption[]{H$\alpha$ profiles for HD 34282 and HD 141569. Spectra of standard
  stars with spectral types A3~V and A0~V, respectively, have been
  subtracted from the original spectra, therefore only the pure emission is
  shown in these panels.  The lines have been shifted vertically by an
  arbitrary amount for clarity, so their wings should be put at the same
  level in order to compare the profiles.  Each profile has been shifted
  horizontally 30 \AA{} with respect to the previous one as indicated in
  the abscissae. The vertical dotted lines mark the rest wavelength for
  H$\alpha$. The spectra have been ordered chronologically from left to
  right (see the dates of the observations in Table \ref{SETUPS}).}
\label{HALPHA}
\end{figure}

\begin{figure}
\epsfig{file=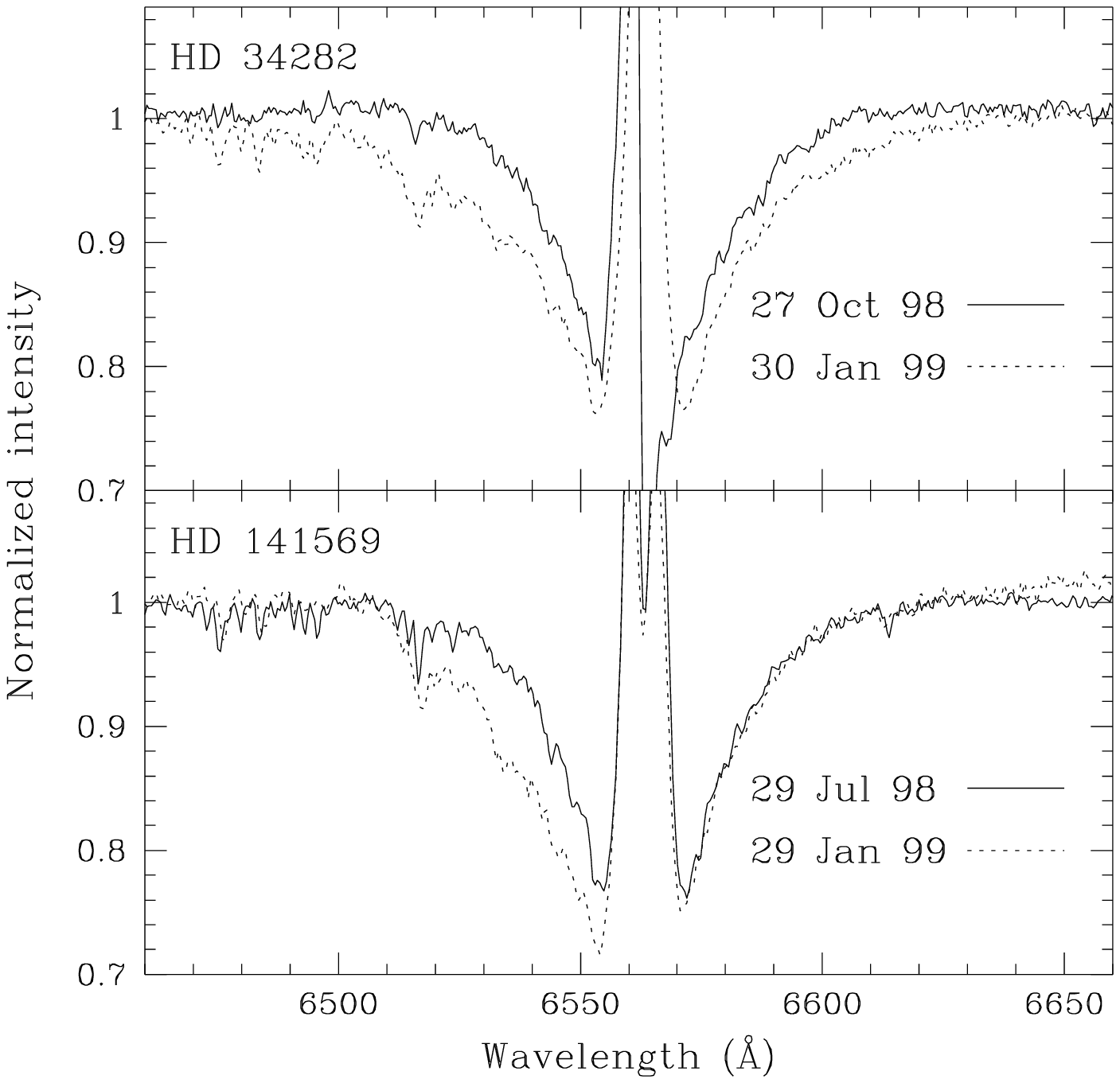,height=9.0cm}
\caption[]{Details of the H$\alpha$ profiles for HD 34282 and HD 141569. 
  The most extreme cases have been taken for each star to illustrate the
  complex behaviour of this line. The variations in the absorption profiles
  appear in both wings but are more pronounced in the blue one.}
\label{HALPHAWING}
\end{figure}

The high and intermediate resolution spectra of both stars show
non-variable early-type photospheric spectra, interstellar (IS) narrow
absorption lines and diffuse IS bands. In both cases, however, the most
prominent features come from the circumstellar (CS) environment. Fig.
\ref{INT} shows the average INT spectra of HD 34282 (January 99) and HD
141569 (May 98) where some lines are identified. In the following, we
describe the CS spectrum of both objects.

{\it HD 34282:} The UES echelle spectra show a CS contribution in the Ca
{\sc ii} K line and in the Balmer lines H$\beta$ and H$\gamma$. The line He
{\sc i} at 5876 \AA ~is seen in absorption; it is unlikely to be of
photospheric origin since HD 34282 has an $\sim$A3 spectral type. Its
strength and profile might show small variations (after a comparison
of all the INT spectra), but this should be confirmed with higher
resolution spectra. He {\sc i} 5876 \AA ~ has been observed with large
variations in HAeBe and T Tauri stars (e.g. B\"ohm \& Catala 1995, Johns \&
Basri 1995) and it has been explained in terms of magnetospheric accretion
in the low-mass PMS stars (e.g. Oliveira et al.  2000), but in the case of
HAeBe stars its origin is controversial (e.g.  Bouret \& Catala 2000).

\begin{table*}
\caption{Compilation of optical ({\it UBVRI}) and near-IR ({\it JHK}) photometry for HD 34282 and HD 141569.} 
\begin{tabular}{lccccccccc}
\hline
Object    & $U$  & $B$  & $V$  & $R$  & $I$  & $J$  & $H$  & $K$  & Refs. \\
\hline
HD 34282  &10.47 &10.23 &10.03 & 9.91 & 9.76 & 9.11 & 8.40 & 7.62 & 1, October 98\\
          &10.19 &10.03 & 9.87 & 9.77 & 9.63 & 9.15 & 8.53 & 7.84 & 1, January 99\\
          &      &      &10.11 &      &      & 9.17 & 8.32 & 7.42 & 2   \\
          &10.16 &10.05 & 9.88 & 9.79 & 9.68 & 9.02 & 8.20 & 7.44 & 3   \\
          &10.15 &10.05 & 9.89 & 9.81 & 9.71 &      &      &      & 4   \\
          &10.11 & 9.97 & 9.81 & 9.73 & 9.62 &      &      &      & 4   \\
          &10.51 &10.07 & 9.94 & 9.95 & 9.79 &      &      &      & 4   \\
          &      &      &      &      &      & 8.99 & 8.54 & 8.08 & 6   \\
\hline
HD 141569 & 7.23 & 7.20 & 7.11 & 7.01 & 7.03 & 6.70 & 6.59 & 6.53 & 1   \\
          &      &      & 7.09 &      &      & 6.88 & 6.86 & 6.82 & 2   \\
          &      &      & 7.13 &      &      & 6.87 & 6.84 & 6.80 & 3   \\
          & 7.24 & 7.21 & 7.10 & 7.03 & 6.96 &      &      &      & 4   \\
          & 7.28 & 7.26 & 7.18 &      &      &      &      &      & 5   \\ 
\hline
\end{tabular}

(1)  EXPORT, (2) Malfait et al. (1998), (3) Sylvester et al. (1996),
(4) de Winter et al. (2001), \\(5) Penprase (1992), (6) Coulson et al. (1998)
\label{TablePHOT}
\end{table*}

H$\alpha$ always presents a variable double-peaked emission with a central
absorption. Fig. \ref{HALPHA} shows the observed H$\alpha$ profiles after
the subtraction of a spectrum of a standard of the same spectral type
broadened to the rotational velocity of HD 34282.  The depth of the central
absorption varies and its radial velocity, measured with respect to the
stellar photosphere, shifts from an average value of 27.5$\pm$5.5 km
s$^{-1}$ in October 98 to --63.3$\pm$14.6 km s$^{-1}$ in January 99. The
radial velocities of the violet and red emission peaks and their separation
also change from October 98 to January 99 (the maximum peak separation is
319 km s$^{-1}$ on 28 Oct 98 and the minimum is 117 km s$^{-1}$ on 30 Jan
99). The relative intensity of the emission peaks changes from day to day
and a flip is observed from October 98, when the violet peak is stronger,
to January 99, when the read peak is the stronger one. Furthermore, the
blue wing shows a shoulder on 28 October 98 which could be due to a third
velocity component. Finally, the total equivalent width varies from --3.0
to --5.7 \AA. Using the $R$ band flux from simultaneous optical photometry
on 27 October 98 and 30 January 99 we estimate a H$\alpha$ line emission
flux of $7.4\times 10^{-13}$ and $1.6\times 10^{-12}$ erg cm$^{-2}$
s$^{-1}$, respectively. The variability in the emission flux and shape of
H$\alpha$ proves the changing conditions in the CS region where this line
originates. Grinin \& Rostopchina (1996) studied a sample of HAeBe stars
and found that double-peaked H$\alpha$ emission likely arises in irregular
gaseous circumstellar disks rotating close to the stars and seen nearly
edge on. We notice that Pi\'etu et al.  (2003) find an inclination of the
CO keplerian disk of $i\!=\!56^\circ$ with respect to the plane of the sky.
The change of the strength ratio between the violet and red peaks could
reflect the relative contribution of the approaching and receding parts of
the disk in the different spectra.  More sophisticated accretion disk
models have recently been developed by Tambovtseva \& Grinin (2000) to
explain the hydrogen lines of HAeBe stars in a context similar to the
widely accepted magnetospheric accretion models for the low-mass T Tauri
stars.

{\it HD 141569:} Double-peaked H$\alpha$ emission is observed in the
spectrum of this star. Such double-peaked emission, though much less
pronounced, is also observed superimposed to the H$\beta$ photospheric
line in all UES spectra. The heliocentric radial velocity of the star
obtained from the UES spectra is --13.1 km s$^{-1}$, which is in good
agreement with previous estimates (Dunkin et al. 1997, Brittain et al.
2003), given the velocity resolution of the UES spectra, 6 km
s$^{-1}$.  The violet and red H$\alpha$ emission peaks have
$-113.2\pm$3.9 km/s and 128.3$\pm$5.0 km/s, respectively, while the
central absorption lies at a velocity of 11.9$\pm$4.2 (average values
and rms error of the 10 INT spectra, velocities measured with respect
to the stellar photosphere). Dunkin et al.  (1997) find agreement
between the stellar and the H$\alpha$ central absorption velocities;
the discrepancy in our values must be treated with caution given the
spectral resolution of the individual INT spectra, $\sim\!50$ km
s$^{-1}$.  The observed velocity separation between both emission
peaks is in good agreement with the results of Dunkin et al.  (1997).
The emission profiles are quite similar in all spectra, although they
vary noticeably.  The violet peak is always stronger than the red one,
but their relative intensity changes (Fig. \ref{HALPHA}); there is
also a variation in the H$\alpha$ blue wing (Fig. \ref{HALPHAWING}).
In addition, the equivalent width varies, ranging from --6.14$\pm$0.03
\AA ~in July 1998, when the violet peak was larger, to --5.63$\pm$0.25
\AA ~in January 1999, when the violet peak was smaller. The average
H$\alpha$ line flux is $2.1 \times 10^{-11}$ erg cm$^{-2}$ s$^{-1}$.
    
All INT spectra show [O {\sc i}] in emission, relatively strong at 6300
\AA{} and very weak at 6364 \AA. The emission is constant and has an
equivalent width W([O~{\sc i}] 6300 \AA) = 0.13 \AA; its flux is $4.6
\times 10^{-13}$ erg cm$^{-2}$ s$^{-1}$. The peak of the emission occurs at
the same radial velocity as the H$\alpha$ absorption peak. [O {\sc i}]
emission is usually interpreted as an indicator of low density winds in
HAeBe stars (B\"ohm \& Catala 1994; Corcoran \& Ray 1997).

\subsection{Photometry and the spectral energy distribution}
\label{PHOT}

\begin{table}[t]
\caption{Absolute dereddened fluxes for the SEDs} 
\begin{tabular}{lr@{.}lcr@{.}l}
\hline
     & \multicolumn{2}{l}{~}             & HD 34282                  &  \multicolumn{2}{c}{HD 141569} \\\hline
Band & \multicolumn{2}{l}{log $\lambda$} & $\log F_\lambda$          & \multicolumn{2}{c}{$\log F_\lambda$}        \\
     & \multicolumn{2}{l}{($\mu$m)}      & (W m$^{-2}\mu{\rm m}^{-1}$)& \multicolumn{2}{c}{(W m$^{-2}\mu{\rm m}^{-1}$)} \\
\hline
{\it IUE} &      --0&86      &    --                  &  --9&90$\pm$0.02 (4) \\
{\it IUE} &      --0&75      &    --                  &  --9&79$\pm$0.02 (4) \\
{\it IUE} &      --0&66      &    --                  &  --9&94$\pm$0.02 (4) \\    
{\it IUE} &      --0&59      &    --                  & --10&01$\pm$0.02 (4) \\
{\it IUE} &      --0&53      &    --                  & --10&08$\pm$0.02 (4) \\  
{\it Wal W}&     --0&49      & --11.45$\pm$0.02 (5) &  \nodata    \\
{\it Wal U}&     --0&44      & --11.44$\pm$0.02 (5) &  \nodata    \\
{\it Wal L}&     --0&42      & --11.23$\pm$0.02 (5) &  \nodata    \\
{\it Wal B}&     --0&37      & --11.11$\pm$0.02 (5) &  \nodata    \\
{\it Wal V}&     --0&26      & --11.32$\pm$0.02 (5) &  \nodata    \\
{\it U}  &       --0&44      & --11.35$\pm$0.01 (1) & --10&02$\pm$0.02 (1)  \\
{\it B}  &       --0&36      & --11.12$\pm$0.01 (1) &  --9&88$\pm$0.02 (1)  \\
{\it V}  &       --0&26      & --11.31$\pm$0.01 (1) & --10&12$\pm$0.02 (1)  \\
{\it R}  &       --0&15      & --11.62$\pm$0.01 (1) & --10&45$\pm$0.02 (1)  \\
{\it I}  &       --0&01      & --11.90$\pm$0.01 (1) & --10&82$\pm$0.04 (1)  \\
{\it J}  &         0&10      & --12.15$\pm$0.02 (1) & --11&22$\pm$0.02 (2)  \\
{\it H}  &         0&21      & --12.32$\pm$0.02 (1) & --11&64$\pm$0.02 (2)  \\
{\it K}  &         0&35      & --12.54$\pm$0.02 (1) & --12&12$\pm$0.02 (2)  \\
{\it ISO} &        0&39      &    --                & --12&27$\pm$0.02 (6)  \\
{\it ISO} &        0&51      &    --                & --12&66$\pm$0.02 (6)  \\
{\it L}   &        0&58      & --12.75$\pm$0.02 (3) & --12&92$\pm$0.02 (3)  \\
{\it ISO} &        0&63      &    --                & --13&14$\pm$0.02 (6)  \\ 
{\it M}   &        0&68      & --13.12$\pm$0.02 (3) & --13&25$\pm$0.02 (3)  \\ 
{\it ISO} &        0&77      &    --                & --13&46$\pm$0.02 (6)  \\  
{\it ISO} &        0&83      &    --                & --13&54$\pm$0.02 (6)  \\  
{\it ISO} &        0&89      &    --                & --13&39$\pm$0.02 (6)  \\  
{\it ISO} &        0&94      &    --                & --13&61$\pm$0.02 (6)  \\  
{\it ISO} &        0&99      &    --                & --13&78$\pm$0.02 (6)  \\  
{\it ISO} &        1&03      &    --                & --13&82$\pm$0.02 (6)  \\  
{\it ISO} &        1&06      &    --                & --13&84$\pm$0.02 (6)  \\  
{\it IRAS} &       1&08      & --13.82$\pm$0.10 (7) & --13&92$\pm$0.05 (7)  \\
{\it IRAS} &       1&40      & --14.09$\pm$0.10 (7) & --14&03$\pm$0.07 (7)  \\
{\it IRAS} &       1&78      & --14.03$\pm$0.10 (7) & --14&32$\pm$0.09 (7)  \\
{\it IRAS} &       2&00      & --14.50$\pm$0.10 (7) & --15&00$\pm$0.10 (7)  \\
0.45 mm &          2&65      & --16.72$\pm$0.23 (3) &  \nodata     \\
0.8 mm  &          2&90      & --17.72$\pm$0.07 (3) &  \nodata     \\
1.1 mm  &          3&04      & --18.34$\pm$0.09 (3) &  $<\!-19$&05 (3)  \\
1.3 mm  &          3&12      & --18.71$\pm$0.09 (8) &  \nodata     \\
1.35 mm &          3&13      &    --                  & --20&05$\pm$0.02 (9)  \\
2.6 mm  &          3&41      & --19.99$\pm$0.13 (10) &  \nodata     \\
3.4 mm  &          3&53      & --20.89$\pm$0.06 (8) &  \nodata     \\\hline
\end{tabular}

(1) EXPORT, (2) Malfait et al. (1998), (3) Sylvester et al. (1996),
(4) INES Database, (5) de Geus et al. (1990), (6) {\it ISO} PHOT-S, TDT No. 62701662,
(7) {\it IRAS} PSC (colour corrected), (8) Pi\'etu et al. (2003),
(9) Sylvester et al. (2001), (10) Mannings \& Sargent (2000)
\label{TableSED}
\end{table}

EXPORT optical {\it UBVRI} and near-IR {\it JHK} magnitudes of HD
32482 and HD 141569 are given in Oudmaijer et al. (2001) and Eiroa et
al.  (2001).  HD 34282 varies in both wavelength regimes from October
98 to January 99, but no variability is detected on time scales of
days. HD 141569 remains constant on all time scales within the
photometric uncertainties (typically a few percents in any photometry
band). Table \ref{TablePHOT} shows the mean magnitudes of HD 34282 as
observed in October 98 and January 99, as well as the mean magnitudes
of HD 141569. For the sake of comparison, photometric results taken
from the literature are also shown. In general, typical intrinsic
error bars in the EXPORT {\it UBVRI} magnitudes are of the order of
$\pm 0.04$ mag and always less than 0.10 mag; for {\it JHK} they are
less than 0.05 mag (see Eiroa et al. 2001 and Oudmaijer et al.  2001
for specific values).

HD 34282 was brighter in the optical in January 99 than in October 98
(the values of $V$ are 10.02, 10.03 and 9.87 for 24, 28 October 98 and
30 January 99, respectively). The January 99 magnitudes likely
correspond to the maximum optical brightness level of this star, as
comparison with other results (Table \ref{TablePHOT}) and with the
{\it Hipparcos} light curve suggests. The near-IR brightness behaves in the
opposite way: the star was brighter in October 98. The optical
variability amplitude decreases with wavelength, as expected from the
usual mechanisms invoked to explain the photometric variability in
PMS stars (e.g. Herbst \& Shevchenko 1999), but in the near-IR the
maximum variability amplitude occurs in $K$. The cause of this
different behaviour might be due to the fact that the optical
variability is due to changes affecting the stellar photosphere while
the near-IR fluxes reflect the contribution of the star and the
circumstellar disk. A simultaneous anticorrelation of the variability
in both wavelength regimes has also been found in other pre-main
sequence stars, though on a much shorter time scale (Eiroa et al.
2002; Grinin et al. 2002).

Concerning HD 141569, our optical photometry and previous results are
in excellent agreement. The apparent discrepancy between our near-IR
magnitudes and those obtained by Sylvester et al. (1996) and Malfait
et al. (1998) is likely due to the fact that the companion star \object{HD
141569B} was at least partly in our beam (Weinberger et al. 2000). Once
our measurements are corrected for this, the agreement is
satisfactory.

For both stars EXPORT photometry together with data from near
ultraviolet to millimetre wavelengths have been used to delineate the
spectral energy distributions, which are, to our knowledge, the most
complete ones to date.  Figs. \ref{hd34282fit} and \ref{hd141569fit}
show the observed spectral energy distributions of both stars. Table
\ref{TableSED} gives dereddened fluxes for each star and corresponding
references (they are also plotted in Figs. \ref{hd34282fit} and
\ref{hd141569fit}).  $E(B\!-\!V)$ values of 0.05 for HD 34282 and 0.12
for HD 141569 (see subsection \ref{THEPARAMS1} for the origin of these
values) were used for the reddening corrections. The extintion curve
of Rieke \& Lebofsky (1985), from 4400 \AA{} to longer wavelengths,
and that of Steenman \& Th\'e (1991) for shorter wavelengths were
used.  We have used the {\it UBVRI} (one set of values) 
and the mean {\it JHK} (three sets) photometry from January 99
for HD 34282 because they likely represent better the stellar
photosphere, i.e. when the star was brighter in the optical. 
For HD 141569, we used the mean EXPORT {\it UBVRI} photometry and {\it JHK}
magnitudes from Malfait et al. (1998).  {\it UBVRIJHK} magnitudes have
been converted into fluxes using the absolute calibration given by
Schmidt-Kaler (1982).  {\it IUE} spectra in the range 1250--3000 \AA
~of HD 141569 are taken from the INES database\footnote{\tt
  http://ines.laeff.esa.es/} (Solano et al. 2001).  No {\it IUE}
observations are available for HD 34282. Walraven photometry for HD
34282 was not contemporaneous with EXPORT observations and has only
been used in subsection \ref{MDOT} for a determination of the disk
mass accretion rate. In the infrared, we have used colour-corrected
{\it IRAS} fluxes and {\it ISO} scientifically validated `Off-Line
Processing' products (OLP) taken from the Data Archive at the {\it
  ISO} Data Centre. Submillimetre and millimetre continuum fluxes have
been taken from the references listed in Table \ref{TableSED}; the
inclusion of the continuum fluxes at these long wavelengths
alleviates, to some extent, the uncertainties in the flux calibration
of the infrared {\it ISO} pipeline-reduced fluxes.

\section{Stellar parameters of the stars and their distances} 
\label{THEPARAMS}

As we mentioned above, both stars have been reported to have anomalous
positions in the HR diagram, falling below the main sequence.  However, the
observational data show that HD 34282 is a highly variable HAeBe star with
an active CS environment.  Concerning HD 141569, though its disk shares
similarities with some debris disk systems, which are found to be located
below the main sequence, the star does show observational properties which
suggest an earlier evolutionary stage. In particular, our data show that
the object retains a non-negligible CS activity, as evidenced by the
[O~{\sc i}] emission and the variable H$\alpha$ emission.  Taking into
account these considerations, both stars should be located above or on the
main sequence, in better agreement with normal expectations from their
photometric and spectroscopic properties, and stellar evolution theory. In
this section we carry out a detailed study of the physical parameters of
the stars leading us to re-consider their location in the HR diagram.
The determination of the effective temperature, gravity and stellar
metallicity will allow us, by using the appropriate set of evolutionary
tracks and isochrones, to estimate the luminosity, mass and age of the
stars, as well as their distances by means of the measured stellar fluxes.

\subsection{Stellar parameters: effective temperature, gravity, 
metallicity and rotational velocity}
\label{THEPARAMS1}

\begin{figure}
\epsfig{file=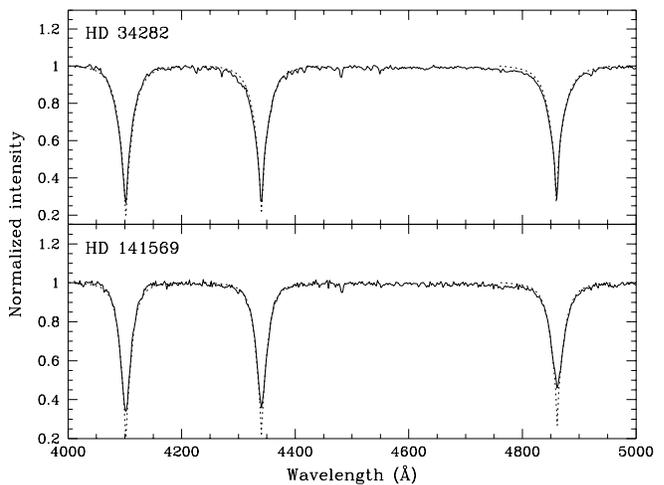,height=7.0cm}
\caption[]{The Balmer lines H$\beta$, H$\gamma$ and H$\delta$ for
  HD 34282 and HD 141569 (solid lines) extracted from the spectra taken
  with CAFOS on the 2.2m telescope at CAHA. The dotted lines are the Kurucz
  synthetic profiles corresponding to the $T_{\rm eff}$, $\log g_*$ and
  metallicity listed in Table \ref{STARS} for each star. The discrepancies
  in the higher parts of the wings arise from contamination with weak metalic lines 
  and also from the difficulties in normalizing the original spectra.}
\label{CAFOS}
\end{figure}

Kurucz (1993) model atmospheres are used to estimate the physical
parameters describing the photospheres of HD 34282 and HD 141569. We
use the following iterative procedure. First, an initial value of the
stellar gravity is obtained from the comparison of synthetic profiles,
computed with solar metallicity, with the observed H$\beta$, H$\gamma$
and H$\delta$ profiles extracted from the CAFOS observations. For this
purpose, a tentative value of $T_{\rm eff}$, based on the spectral
type of each star, is taken. The gravity is obtained by measuring the
width of the observed Balmer profiles at one intensity level below the
normalized continuum, namely $I\!=\!0.80$; this has been chosen to
avoid potential artifacts in the wings near the continuum arising from
the normalization of the original spectra. The widths of the Balmer
lines are then compared with those measured on a grid of synthetic
profiles computed for different gravities\footnote{The convolution of
  the theoretical profiles with the instrumental response changes the
  results by a minute quantity.  For HD 141569, which is the star with
  less intrinsic uncertainties in the determination of the gravity,
  the relative difference with the value in Table \ref{STARS} is  about
  0.5\%.}. A linear interpolation is done taking the widths of each
Balmer line as independent variables and the gravity as the unknown.
The result for $\log g_*$ is the mean value of all the interpolations
with an uncertainty corresponding to the standard deviation.  

With this value of $\log g_*$ fixed, we re-estimate the efective
temperatures of the stars and their metallicities. This is done by means of
a systematic comparison of synthetic line profiles, obtained using Kurucz's
models of different temperatures and metallicities, with the high
resolution WHT/UES spectra. The regions of effective temperatures from 8500
to 9750 K for HD 34282, and from 9750 to 10500 K for HD 141569 were
explored, which approximately encompass the values corresponding to the
spectral types reported for both stars. For each temperature, different
metallicities were used to synthesize spectral lines in several selected
spectral intervals. The synthetic spectra were rotationally broadened with
the corresponding $v\sin i$ values for each star (Mora et al.
2001).  A microturbulence velocity of 2 km/s was used throughout.

For HD 34282 the lines chosen are: 4404.75, 4957.60 \AA{} (Fe {\sc
  i}), 4508.29, 4515.34, 4583.84, 4629.34 \AA{} (Fe {\sc ii}) and
4501.27 \AA{} (Ti {\sc ii}). For HD 141569 the projected rotational
velocity is so high that it is difficult to find individual metallic
lines; instead the regions 4260--4280 \AA, 4375--4400 \AA{} and
4540--4560 \AA{}, which contain a large number of lines, have been
selected.  For both stars the lines 3933.66 \AA{} (Ca {\sc ii}),
4481.13 \AA (Mg {\sc ii}), and the Si {\sc ii} doublet at 6347.11,
6371.37 \AA{} have also been used (the Si {\sc ii} doublet is clearly
observed in the INT spectra).  It is found, particularly in the case
of HD 34282, that different ($T_{\rm eff}$, [Fe/H]) pairs fit the
observed spectra reasonably well.  The comparison shows that higher
effective temperatures require higher metallicities to fit a given
spectral line.  This ($T_{\rm eff}$, [Fe/H]) degeneracy can be solved
by using the observed SEDs, in particular the {\it UBVRI} photometry.

For each ($T_{\rm eff}$, [Fe/H]) pair, which provides an apparent
satisfactory fit to the observed spectra, we have computed the
corresponding synthetic Kurucz SED in the spectral interval from 3000
to 10000 \AA, i.e., the range from $U$ to $I$. The observed {\it
  UBVRI} fluxes from each star have been dereddened at small steps of
$E(B\!-\!V)$ and compared to the Kurucz synthetic models (we have
covered the interval $0.0 < E(B\!-\!V) < 0.30$ with a stepsize of
0.01). The observed and synthetic photometry were normalized to $R$.
It turns out that the overall SED shape is significantly more
sensitive to the effective temperature than to the metallicity; thus,
this method is particularly useful to pinpoint the best $T_{\rm eff}$,
and consequently the ($T_{\rm eff}$, [Fe/H]) solution.  The best
$T_{\rm eff}$ value fitting the data is chosen by means of a minimum
$\chi^2$ criteria. Once the best values for $T_{\rm eff}$ and [Fe/H]
have been found, the value of the gravity is re-calculated using the
procedure described above with synthetic Balmer profiles corresponding
to this metallicity. The process is iterated until convergence is
achieved.
 
\begin{figure*}
\epsfig{file=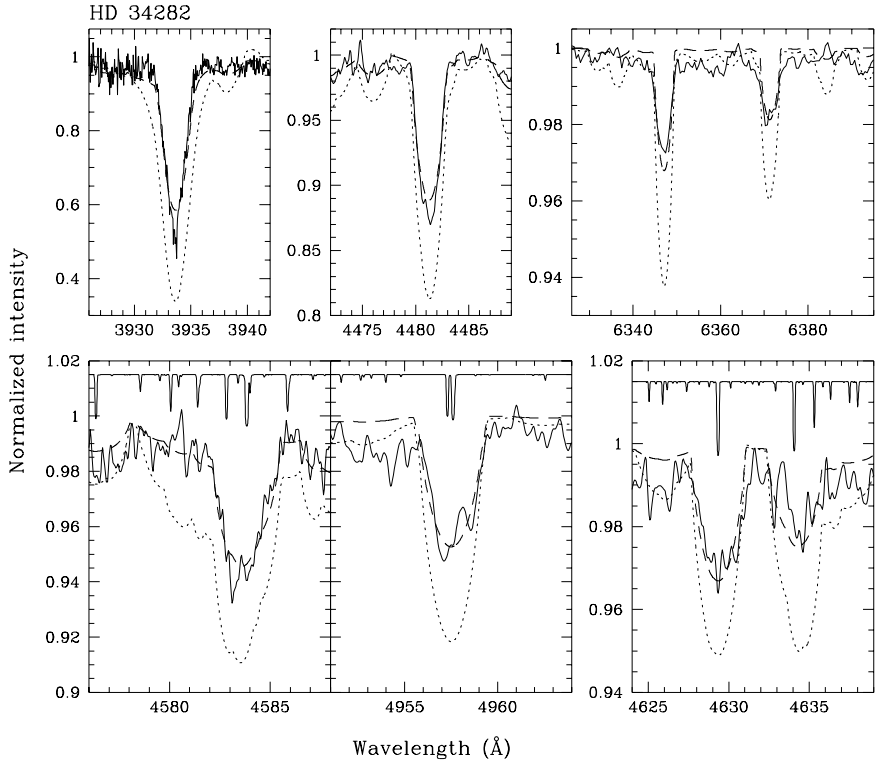,height=11.0cm}
\caption[]{Comparison of observed and synthetic spectral features of HD 34282.
  From top to bottom and left to right the following lines have been
  plotted: Ca {\sc ii} K (3933.66 \AA), Mg {\sc ii} (4481.13 \AA), Si {\sc
    ii} (6347.11, 6371.37 \AA), Fe {\sc ii} (4583.84 \AA), Fe {\sc i}
  (4957.60 \AA) and Fe {\sc ii} (4629.34 \AA).  Observed spectra are
  plotted as solid lines. Synthetic spectra computed with $T_{\rm
    eff}\!=\!8625$ K, [Fe/H]=--0.8 and $\log g_*\!=4.20$ and broadened with
  $v \sin i\!=\!110$ km/s are plotted as dashed lines.  Dotted lines
  represent synthetic spectra computed with the same $T_{\rm eff}$, $\log
  g_*$ and $v \sin i$ but using solar abundances. In the lower panels, the
  synthetic unbroadened spectra are shown to illustrate the lines
  giving rise to the observed profiles.}
\label{HD34282metals}
\end{figure*}

\begin{figure*}
\epsfig{file=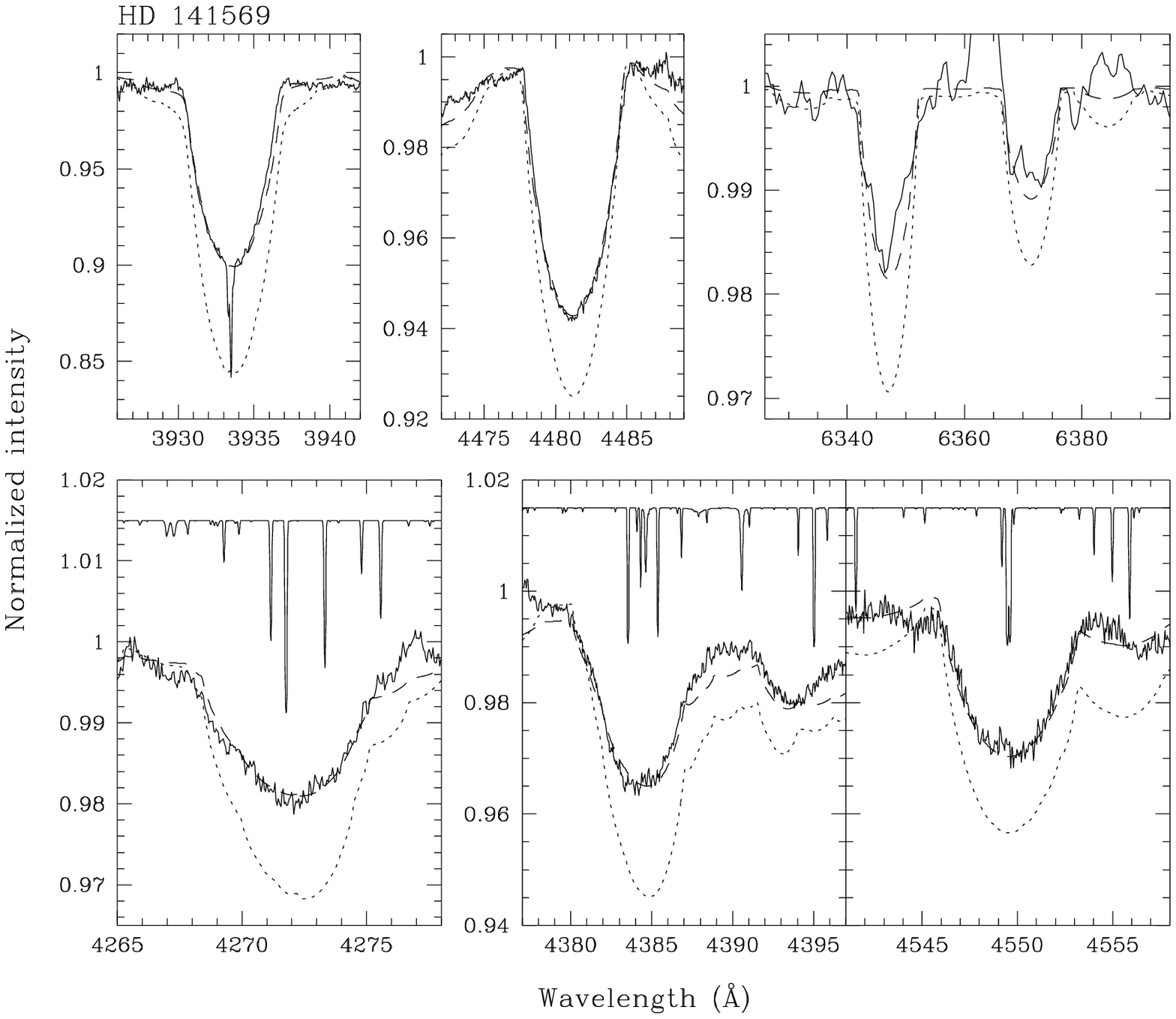,height=11.0cm}
\caption[]{Comparison of observed and synthetic spectral features of
  HD 141569. In the top panels, the spectral lines are the same as in
  Fig.  \ref{HD34282metals}.  In the bottom panels, three spectral
  regions, instead of individual lines, are displayed. Observed
  spectra are plotted as solid lines. Synthetic spectra computed with
  $T_{\rm eff}\!=\!10000$ K, [Fe/H]=--0.5 and $\log g_*\!=4.28$ and
  broadened with $v \sin i\!=\!236$ km/s are plotted as dashed lines.
  Dotted lines represent synthetic spectra computed with the same
  $T_{\rm eff}$, $\log g_*$ and $v \sin i$ but using solar abundances.
  In the lower panels, the synthetic unbroadened spectra are shown to
  illustrate the lines giving rise to the observed profiles.}
\label{HD141569metals}
\end{figure*}

The optimum solutions found are (see also Table \ref{STARS}): HD
34282, $T_{\rm eff}$ = 8625 K, [Fe/H] =$-0.8$, $\log g_*$ = 4.20; HD
141569, $T_{\rm eff}$ = 10000 K, [Fe/H] =$-0.5$, $\log g_*$ = 4.28.
The corresponding spectral types for these temperatures and gravities
are A3~V and B9.5~V respectively.  The relevant result is that both
stars have {\it low} metal abundances.  In addition, projected
rotational velocities, $v\sin i$, of 110 km s$^{-1}$ for HD 34282 and
236 km s$^{-1}$ for HD 141569 provide a better agreement with the
observed profiles than the $v \sin i$ values reported by Mora et al.
(2001). Finally, the best $\chi^2$ fit gives $E(B-V)$ = 0.05 mag for
HD 34282 and 0.12 mag for HD 141569. We note that Mora et al.  (2001)
deduced $E(B\!-\!V)=0.15 \pm 0.10$ mag for the latter star from the
graphite interstellar absorption band at 2200 \AA{} observed with the
{\it IUE}, also Oudmaijer et al. (2001) found $A_V=0.34\pm 0.20$,
which implies $E(B\!-\!V)=0.11 \pm 0.06$; both values of the
extinction are in very good agreement with our present estimate.
Figs.  \ref{HD34282metals} and \ref{HD141569metals} show the
comparison between the observed and best synthetic spectra for several
spectral lines and ranges; synthetic spectra using solar abundances
are also plotted for comparison.  Both figures clearly show that the
observed spectra of HD 34282 and HD 141569 are much better reproduced
with our low metallicity estimates and, therefore, a low metal content
in the photospheres of both stars is strongly suggested. We note that
HD 34282 has some similarities with \object{HD 37411}, another low
metallicity PMS star (Gray \& Corbally 1998).  With respect to HD
141569, Gray \& Corbally (1998) indicate that it is a `mildy
metal-weak' star, in agreement with our results, though Dunkin et al.
(1997) give a solar abundance.

\begin{table}[h]
  \caption[]{Stellar parameters determined.}
   \begin{tabular}{lll}
                      &\multicolumn{2}{c}{HD 34282}                             \\\hline
Effective temperature &\multicolumn{2}{c}{8625$\pm 200$ K}                      \\
Spectral type         &\multicolumn{2}{c}{A3 V}                                 \\
$v \sin i$            &\multicolumn{2}{c}{$110\pm 10$ km/s}                     \\
Metallicity           &\multicolumn{2}{c}{[Fe/H]=$-0.8\pm 0.1$}                 \\
Log $g_*$             &\multicolumn{2}{c}{4.20$\pm 0.20$}                       \\
$E(B-V)$              &\multicolumn{2}{c}{0.05}                                 \\\hline
Stellar flux ($F_*$)  &\multicolumn{2}{c}{$(3.61 \pm 0.44)\times10^{-12}$ W m$^{-2}$}  \\
IR Excess ($F_{\rm IR}$)&\multicolumn{2}{c}{$(2.28 \pm 0.40)\times10^{-12}$ W m$^{-2}$}\\
$\lambda_{\rm IR}$    &\multicolumn{2}{c}{1.25 $\mu$m}                           \\\hline  
                      &[$\pm\Delta\log g_*$]       & [$\pm\Delta T_{\rm eff}$]  \\\hline
$L/L_\odot$           &13.64$^{-5.36}_{+12.02}$    &13.64$^{+1.78}_{-1.53}$     \\
Mass (M$_\odot$)      &1.59$^{-0.07}_{+0.30}$      &1.59$^{+0.04}_{-0.05}$      \\
Age (Myr)             &6.41$^{+1.92}_{-2.58}$      &6.41$^{-0.41}_{+0.46}$      \\
Distance (pc)         &348$^{-77}_{+129}$          &348$^{+22}_{-20}$           \\\hline
                      &                            &                            \\
                      &\multicolumn{2}{c}{HD 141569}                            \\\hline
Effective temperature &\multicolumn{2}{c}{10000$\pm 200$ K}                     \\
Spectral type         &\multicolumn{2}{c}{B9.5 V}                               \\
$v \sin i$            &\multicolumn{2}{c}{$236\pm 15$ km/s}                     \\
Metallicity           &\multicolumn{2}{c}{[Fe/H]=$-0.5\pm 0.1$}                 \\
Log $g_*$             &\multicolumn{2}{c}{4.28$\pm 0.04$}                       \\
$E(B-V)$              &\multicolumn{2}{c}{0.12}                                 \\\hline
Stellar flux ($F_*$)  &\multicolumn{2}{c}{$(7.16 \pm 0.24)\times10^{-11}$ W m$^{-2}$}  \\
IR Excess ($F_{\rm IR}$)&\multicolumn{2}{c}{$(1.27 \pm 0.27)\times10^{-12}$ W m$^{-2}$}\\
$\lambda_{\rm IR}$    &\multicolumn{2}{c}{3.25 $\mu$m}                           \\\hline  

                      &[$\pm\Delta\log g_*$]       & [$\pm\Delta T_{\rm eff}$]  \\\hline
$L/L_\odot$           &25.77$^{-2.20}_{+1.63}$     &25.77$^{+3.01}_{-2.54}$     \\
Mass (M$_\odot$)      &2.00$^{+0.01}_{-0.01}$      &2.00$^{+0.06}_{-0.05}$      \\
Age (Myr)             &4.71$^{+0.09}_{-0.15}$      &4.71$^{-0.31}_{+0.33}$      \\
Distance (pc)         &108$^{-5}_{+3}$             &108$^{+6}_{-6}$             \\\hline
   \end{tabular}
\label{STARS}
\end{table}

\subsection{Stellar fluxes and infrared excesses. Luminosities, masses, 
ages and distances.}
\label{STELLARFLUXES}

Once the stellar parameters defining the photospheres of HD 34282 and HD 
141569  have been determined, we can calculate the stellar fluxes and 
infrared excesses. In addition, we can locate the stars in a HR diagram using
appropriate evolutionary tracks and estimate their distances.
 
The stellar flux, $F_*$, is computed by integrating the best-fit
Kurucz model normalized to the dereddened observed $R$-band flux, as
given in Table \ref{TableSED}. The infrared excess, $F_{\rm IR}$, is
calculated by subtracting the flux under the tail of the best-fit
Kurucz model from the observed fluxes, and integrating the result from
$\lambda_{\rm IR}$, the wavelength where the SED separates from the
tail of the Kurucz model (see Figs. \ref{hd34282fit} and
\ref{hd141569fit}). Extrapolations to long wavelengths following the
formula by Chavarria et al. (1981) have been included in the
integration. Table \ref{STARS} shows the integrated fluxes for the
stars and their IR excesses. HD 34282 has a $F_{\rm IR}$/$F_*$ ratio
of 0.63 $\pm$ 0.19, which suggests that some intrinsic luminosity
might be arising from the disk; $F_{\rm IR}$/$F_*$ values of
$\sim$0.25 and $\sim$0.5 are expected from flat and flared optically
thick reprocessing disks, respectively (Kenyon \& Hartmann 1987).  In
the case of HD 141569, we find $F_{\rm IR}$/$F_*$ = 0.018 $\pm$ 0.004
and a passive disk with relatively low optical depth is
suggested. This figure is one order of magnitude higher than typical
values in Vega-type stars, (10$^{-5}$ to 10$^{-3}$, Sylvester et
al. 1996). We note that the ratio $F_{\rm IR}$/$F_*$ = $8 \times
10^{-3}$ given by Zuckerman et al.  (1995) and usually quoted in the
literature does not include the 1.35 mm flux measured by Sylvester et
al.  (2001). If we remove that flux from our estimates, we recover the
figure given by Zuckerman et al.

\begin{figure*}
\begin{center}
\epsfig{file=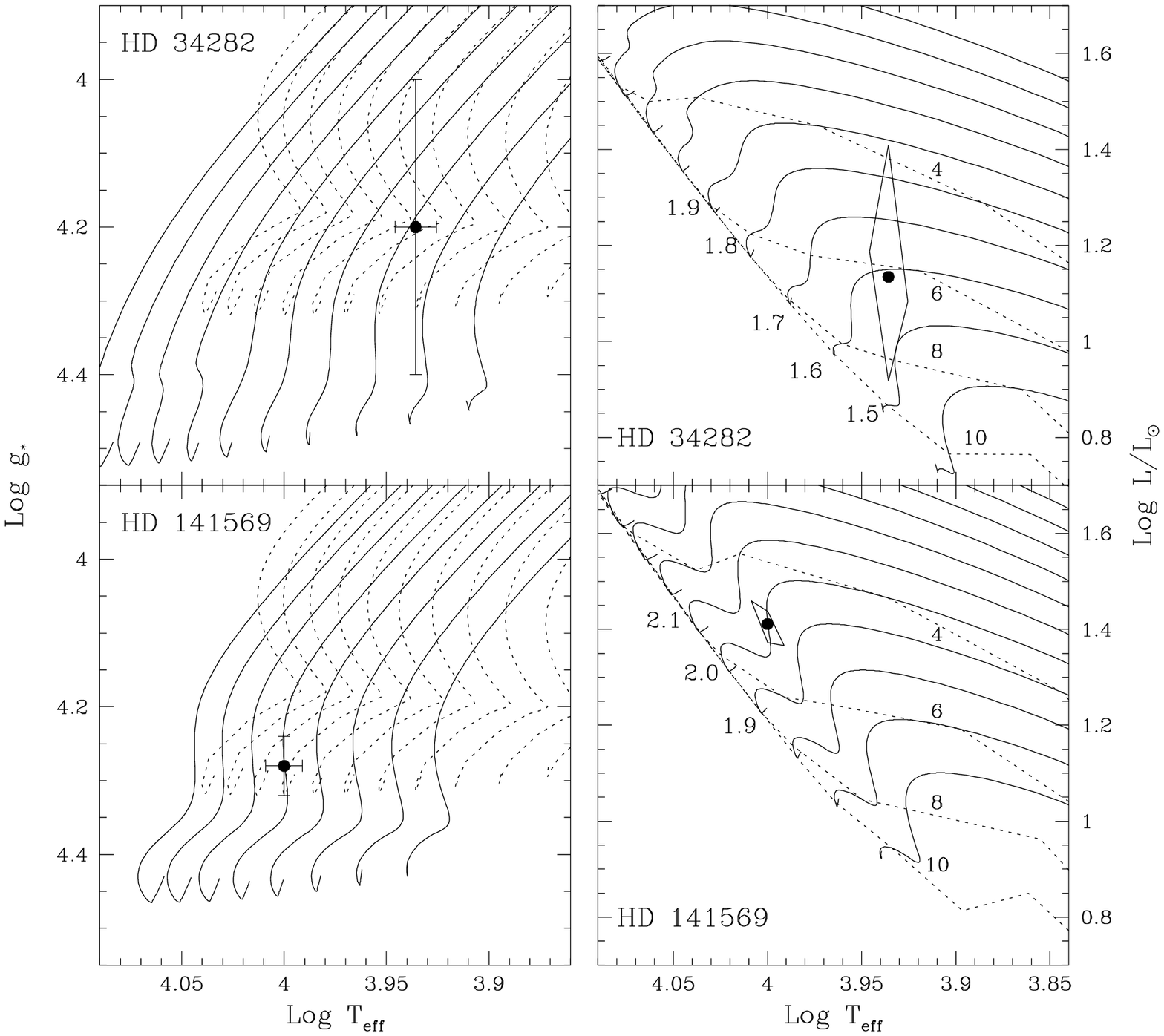,height=12.0cm}
\caption[]{Positions of the stars in the HR diagram according to the results 
  shown in Table \ref{STARS}. {\bf Left.} Upper panel: Position of HD 34282
  in the HR diagram $\log g_*$ -- $\log T_{\rm eff}$. The Y$^2$ PMS tracks
  with $Z\!=\!0.004$ (solid lines) and $Z\!=\!0.02$ (solar, dotted lines)
  are plotted. Lower panel: same as in the upper panel for HD 141569.  The
  tracks correspond to $Z\!=\!0.007$ (solid lines) and $Z\!=\!0.02$ (dotted
  lines). {\bf Right.} Upper panel: Position of HD 34282 in the HR diagram
  $\log L/L_\odot$ -- $\log T_{\rm eff}$. The PMS tracks are those for
  $Z\!=\!0.004$ (solid lines). Lower panel: same as in the upper panel for
  HD 141569. The tracks correspond to $Z\!=\!0.007$ (solid lines). In both
  cases we have labelled the tracks with the corresponding masses and the
  isochrones for 4, 6, 8 and 10 Myr are shown with dotted lines.}
\label{HRdiagrams}
\end{center}
\end{figure*}  

Knowledge of the stellar metallicities is relevant since changes in the
metal abundance, $Z$, produce non-negligible changes in the relative
positions of the evolutionary tracks. In particular, a decrease in the
metallicity moves one given track with a fixed value of the stellar mass,
to lower luminosities, higher gravities and higher temperatures, apart from
changing its overall shape. This effect is illustrated in Fig.
\ref{HRdiagrams} for the $\log g_*$ -- $\log T_{\rm eff}$ HR diagram.  The
left two panels of Fig. \ref{HRdiagrams} show the positions of the stars in
the HR diagram according to the results found for $T_{\rm eff}$ and $\log
g_*$. The evolutionary tracks for a scaled solar mixture from the
Yonsei-Yale group (Yi et al.  2001) -- Y$^2$ in their notation -- have been
used in this work.  The Y$^2$ tracks with $Z\!=\!0.004$ and $Z\!=\!0.007$
are appropriate for HD 34282 and HD 141569, respectively. For comparison,
the tracks with solar metallicity ($Z\!=\!0.02$) are also included in the
plot\footnote{In the original Y$^2$ set, there are no tracks for
  $Z\!=\!0.003$ and 0.006, the appropriate values for each star
  considering that the relationship Fe/Fe$_\odot$=X/X$_\odot$ holds
  (in number if particles), where X means any element with atomic number
  larger than 3.  The closest sets of tracks are those with $Z\!=\!0.004$
  and 0.007 respectively.  We have preferred to use original tracks kindly
  provided by the authors than interpolating between adjacent grids to
  build new sets of tracks and isochrones. In any case the differences are
  expected to be very small.}.

The corresponding values of the effective temperature and gravity are
compared with the set of tracks and values for the luminosity, mass
and age are derived.  The position of HD 34282 and HD 141569 in the HR
diagram $\log L/L_\odot$ -- $\log T_{\rm eff}$ are plotted in the two
panels at the right hand side of Fig. \ref{HRdiagrams}. The
luminosities, combined with the total stellar fluxes $F_*$ obtained,
imply distances of 348 pc for HD 34282, and 108 pc for HD 141569.  All
the results are summarized in Table \ref{STARS}. The results for the
luminosity, mass, age and distance are given in two columns for each
star: in the first one the error bars derive from the uncertainties in
$\log g_*$, keeping $T_{\rm eff}$ fixed; in the second, the error bars
derive from the uncertainties in $T_{\rm eff}$, keeping $\log g_*$
fixed. The uncertainties for HD~34282 are larger than those for
HD~141569 because of the larger uncertainties in the determination of
the gravity. The vertices of the `diamond-like' boxes appearing in the
two right-hand panels of Fig. \ref{HRdiagrams} correspond to the tips
of the error bars in the panels on the left.

Fig. \ref{HRdiagrams} shows that both stars lie above the main
sequence in positions in agreement with their observational properties
and current PMS evolution theory. The {\it anomalous positions} in the
HR diagram previously reported for these stars referred to standard
sets of tracks or photometric parameters computed with solar
metallicity and a very doubtful distance in the case of HD 34282. For
this star the anomalous position in the HR diagram is resolved both by
the low metallicity and an appropriate distance estimate. In the case
of HD 141569, its anomalous position originated only from the use of the
wrong metallicity in the definition of the main sequence\footnote{Note
  that in the two left-hand-side diagrams of Fig. \ref{HRdiagrams}
  both stars also occupy correct positions in the HR diagram when
  compared with tracks with solar metallicity. If these tracks were
  used to obtain values of masses, luminosities and ages, these
  quantities would be incorrect.}.

The distance we obtain for HD 34282 (348$^{+129}_{-77}$ pc) is
comparable to that given by Pi\'etu et al.  (2003)
(400$^{+170}_{-100}$ pc).  Both values are remarkably similar, and are
more than $2\sigma$ larger than the {\it Hipparcos} value. This leads us to
consider that the {\it Hipparcos} parallax for this object has been
overestimated. The stellar parameters quoted by Pi\'etu et al. (2003)
are compatible with ours, although we note their larger errors and
their use of indirect methods to estimate the temperature and spectral
type of the star; in addition, they use solar metallicity tracks which
are not appropriate for this star.  In the case of HD 141569 our
distance is comparable to the {\it Hipparcos} one (99$^{+9}_{-8}$ pc).
We note that the age estimate by Weinberger et al.  (2000) for the two
nearby low mass stars is similar to that deduced by us for HD 141569
directly.  This result reinforces the idea that the three stars are
real companions.

Finally, estimates of the extinctions and distances can also be made
using the absolute magnitudes and colours provided by the Y$^2$
isochrones. Taking the values of mass, age and
metallicity given in Table \ref{STARS} and the observed values of $B$
and $V$ used to construct the SEDs (see subsection \ref{PHOT} and
Table \ref{TablePHOT}), the extinctions $E(B\!-\!V)$ obtained are 0.04
for HD 34282 and 0.06 for HD 141569. The former value especially is in
excellent agreement with that given by the $\chi^2$ analysis. The
absolute magnitudes $M_{\rm V}$ provided by the tracks are 2.04 and
1.37 respectively; these, combined with the observed values of $V$
corrected with the above extinctions, yield distances of 347 pc for
HD 34282 and 129 pc for HD 141569, again in good agreement with
the results shown in Table \ref{STARS}.

\section{Modelling the SEDs: the irradiated accretion disk models}
\label{THESEDS}

The origin of the IR excess in HAeBe stars has been a matter of
intense debate (see e.g. Waters \& Waelkens 1998). Some authors argued
in favor of models with an approximately spherical dusty envelope of
low optical depth (Berrilli et al. 1992; di Francesco et al. 1994;
Pezzuto et al. 1997; Miroshnichenko et al. 1997), others considered
circumstellar disks (Hillenbrand et al. 1992; Chiang et al. 2001;
Natta et al. 2001) or proposed the existence of both (Natta et al.
1993; Miroshnichenko et al. 1999).  Ultimately the existence of disks
around HAeBe stars was firmly established by direct imaging at
millimetre wavelengths (Mannings et al. 1997; Mannings \& Sargent
1997, 2000; Testi et al. 2001; Pi\'etu et al. 2003), in the optical
(Grady et al. 1999, 2000) and in the IR (Close et al. 1997;
Jayawardhana et al.  1998; Weinberger et al. 1999). Some authors have
been able to fit the complete SEDs of HAeBe stars with passive
irradiated disk models from Chiang \& Goldreich (Chiang et al. 2001;
Natta et al.  2001).

\subsection{The models}

In this work we use the self-consistent irradiated accretion disk
models from D'Alessio et al. (1998, 1999 and 2001) to fit the SEDs of
HD 34282 and HD 141569. The methods used here were applied succesfully
to fit the SEDs of young T Tauri stars with an approximate age of 1
Myr. This approach is valid for these two stars since they are
young (see Section \ref{THEPARAMS}) and at least one of them seems to
be still actively accreting. Since the calculations yield the vertical
structure and emission properties of the disk self-consistently with
the stellar parameters, and we have characterized the stellar
photospheres with precision, a good fit to the SED would provide a
physically based picture of the system without the use of {\it ad-hoc}
parametrizations for the disk temperature or surface density profiles.

The models involve the following assumptions: the disk is in a steady
state ($ \dot{M}\!={\rm d}M\!/\!{\rm d}t$ is constant), it is
geometrically thin ($H/R\!<<\!1$, where $H$ is the scale height of the
disk and $R$ is the radial distance), the accretion viscosity is
computed with $\nu\!=\!\alpha\,H\,c_{\rm s}$, following the
$\alpha$-prescription from Shakura \& Sunyaev (1973), in which $c_{\rm
  s}$ is the sound speed. The dust and gas are well mixed in the whole
disk with the usual dust to gas mass ratio of 1/100. For the dust we
use a grain size distribution $n(a)\,=\,n_0\,a^{-p}$ with $p$ values
of 2.5 or 3.5 and a minimum grain size of 0.005 $\mu$m. Possible
values for the maximum grain size are 1 $\mu$m to 10 cm. The
abundances are those given by Pollack et al. (1994). The radiation
field is considered in two separate regimes (one characteristic of the
disk local temperature and one characteristic of the stellar effective
temperature) as in Calvet et al. (1991, 1992) and finally the
radiative transfer is done by solving the first two moments of the
radiative transfer equation with the Eddington approximation (see
D'Alessio et al. 1998, 1999 and 2001 for further details).

The disk is the solution of the detailed vertical structure equations
for each annulus, and the synthetic SEDs are obtained by solving the
radiative transport equation in rays parallel to the line of sight and
integrating (in solid angle) the disk's emission.

Note that, although the models we have applied are amongst the most
sophisticated available, they have limitations inherent in the
complexity of the physics and the computational techniques. For
example, they assume a {\it unique} dust size distribution with a
maximum grain size $a_{\rm max}$ and the same viscosity parameter
$\alpha$ throughout, which in some real cases precludes fitting the
observed SED with a {\it single} component. For HD 34282 a combination
of two disk models is needed: one for the long wavelength part of the
SED, with high viscosity and a maximum grain size of 1 cm (subsection
\ref{LONGWAVE}) and a second one for the mid-infrared region of the
SED, with lower viscosity and a maximum grain size of 1 $\mu$m
(subsection \ref{MIDIR}); the first one will be referred to as the
{\it mid-plane disk} and the latter as the {\it surface disk}
according to their vertical scale heights, of which the first has a
lower value.  We found that these two models were the only combination
able to reproduce the whole SED of this star, implying that some
vertical grain segregation may be present in the disk around HD 34282.
In addition, to explain the bump at 3 $\mu$m we propose the presence
of a {\it wall-like} structure at the inner edge of the disk
(subsection \ref{NEARIR}). For HD 141569 a single disk model accounts
for the whole SED (subsection \ref{MODELHD141569}).

\subsection{The disk model for HD 34282}
\label{MODELHD34282}

The disk around HD 34282 shows signs of activity: optical and near-IR
variability, H$\alpha$ variable emission and high fractional IR
excess.  Hence, to fit the SED of HD 34282, we assume that the disk is
heated by irradiation from the central star and viscous dissipation
from the mass accretion onto the star. 

For the central star the following stellar parameters are assumed
(Table \ref{STARS}): $T_{\rm eff}\!=\!8625$ K, $R_*\!=\!1.66$
R$_\odot$, $M_*\!=\!1.59$ M$_\odot$, $L_*\!=\!13.64$ L$_\odot$ and
$d\!=\!348$ pc, where the stellar radius was calculated from
the stellar mass and the surface gravity. For the disk we
take an inclination angle of $i\!=\!56^\circ$ (Pi\'etu et al. 2003)
and an outer radius of 705 AU, which is the disk radius given by these
authors scaled from their distance of 400 pc to our value of 348
pc. The inner disk radius is calculated self-consistently using the
position of the near-IR bump, the result being 0.31 AU (see subsection
\ref{NEARIR}).

To find the remaining input disk parameters we need to know the mass
accretion rate towards the star (which will be derived from ultraviolet
Walraven photometry), the disk viscosity $\alpha$ (which will be fitted to
raise the appropriate flux at millimetre wavelengths once we know the mass
accretion rate) and the dust properties (related to the slope in the
millimetre range and the IR excess shape at shorter wavelengths). 

\subsubsection{The mass accretion rate in the disk of HD~34282}
\label{MDOT}

The fractional IR excess, $F_{\rm IR}/F_*$, of HD 34282, namely 0.63
$\pm$ 0.19 (Table \ref{STARS}), is a little bit larger than the values
measured in the SEDs of the HAeBe stars studied by Natta et
al. (2001), all of them being smaller than 0.45. These authors reproduce the
SEDs assuming that no accretion is present, the reprocessing of
radiation being the mechanism invoked to explain the IR excess.  We propose
that, given the more prominent IR excess of this star compared to
those of the HAeBe stars in Natta et al. (2001) and the higher
fractional IR excess (slightly larger than 0.5), accretion may be
present in this object. For this reason, we first try to estimate the
mass accretion rate towards HD 34282.

This parameter has been studied using Walraven ultraviolet
photometry of HD 34282 (de Geus et al. 1990) and shock emission models
by Muzerolle et al. (2003a), who apply to HAeBe stars the
magnetospheric accretion scenario from Calvet \& Gullbring (1998) and
Gullbring et al. (2000). This theory explains the UV excess emission
seen in many T Tauri stars as arising from an accretion shock. In
HAeBe stars, the emission from the shocked atmosphere modifies the
underlying photospheric emission reducing the Balmer jump with respect
to the normal main-sequence value in cases with high mass accretion
rate and large filling factors.

\begin{figure}
\epsfig{file=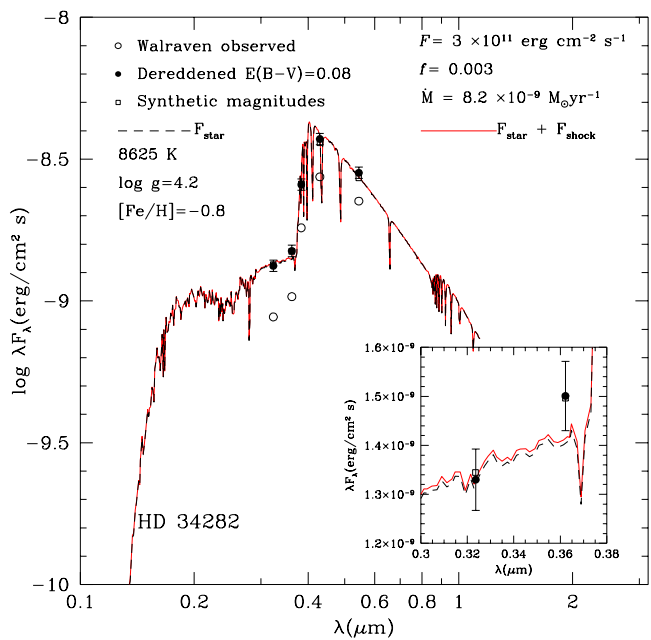, height=9.0cm}
\caption[]{Best-fit shock model fitting to the Walraven photometry of
  HD 34282. The open and filled circles are observed and dereddened
  Walraven photometry, respectively. The open squares are synthetic
  photometry computed with the total (stellar plus shock) flux.  The
  solid line is the Kurucz model plus the accretion shock flux for the
  accretion column energy flux {\it F} and filling factors {\it f}
  specified in the plot. Shifted a small amount below this line would
  lie the Kurucz model for the stellar parameters of the star. The
  magnitude of this tiny difference can be seen in the inset where the
  Kurucz model has been plotted with a dashed line.  The photometric
  point at 0.36 $\mu$m lies above the models because the Walraven $B$
  filter includes fluxes on both sides of the Balmer jump, hence the
  average is higher than the local flux at the blue side of the jump.
  The agreement between the dereddened and synthetic photometry is
  compatible with the presence of a weak accretion shock signature in
  the ultraviolet colours of the star.}
\label{hd34282mdot}
\end{figure}  

Shock models for a central star with the HD 34282 stellar parameters
were computed. The energy flux of the accretion column, {\it F}, is a
free parameter that controls the SED emitted by the accretion shock
and the total mass accretion rate is proportional to this energy flux
and to the filling factor, {\it f}, for the accretion column  (Muzerolle 
et al. 2003a).

Fig. \ref{hd34282mdot} shows the fit of the photospheric plus
accretion flux (solid line) to the dereddened Walraven photometry.
Synthetic Walraven fluxes have been estimated by convolving the total
flux (Kurucz plus shock fluxes) with the Walraven passbands, and
compared to the dereddened data. These synthetic points are plotted as
squares in Fig. \ref{hd34282mdot} and match the dereddened data
(filled circles) so well that they are barely visible in the graph.
In the inset we display an enlargement, in linear scales, of the
region to the blue side of the Balmer jump to show the subtle
difference between the stellar (dashed line) and stellar plus shock
(solid line) fluxes.

We used the $\chi^2$-minimization method to check if some UV excess
was present in HD 34282. For three typical values of the column energy
fluxes (namely $1\times10^{11}$, $3\times10^{11}$ and $1\times10^{12}$
erg cm$^{-2}$ s$^{-1}$) we added the accretion flux to the Kurucz
model with different filling factors {\it f} and performed the
$\chi^2$ estimation for the fittings. Since the Walraven photometry
was not contemporaneous with the EXPORT observations, the best-fit
value for $E(B\!-\!V)$ was 0.08 in this case. The quality of the fits
to the data is very good, but, as the observational errors are much
larger than the difference between models with and without accretion
flux, we can not distinguish between these models. This happened for
all three values of the column energy flux used and in all cases the
mass accretion rate derived was very similar, the average value being
$\dot{M}\!=\!(8.2\pm0.2)\times10^{-9}$ M$_\odot$ yr$^{-1}$. This value
is consistent with the low accretion rates expected in HAeBe stars
(Hartmann et al. 1993; Ghandour et al. 1994; Natta et al.  2001) and
also with estimates for UX Ori-like stars (Tambovtseva \& Grinin
2000).

Given the tiny difference introduced by the shock and the small $\chi^2$
values obtained, it is clear that the excess flux is virtually impossible
to measure in this case and that this determination constrains the mass
accretion rate mostly as an upper limit. Once this method has been applied
to other HAeBe stars with different amounts of IR excess, we will have
better knowledge of its reliability. However, here we will adopt this value
for the mass accretion rate toward the star because, being compatible with
the ultraviolet photometry, it is also consistent with the physical
parameters needed to fit the SED millimetre fluxes (see section
\ref{LONGWAVE}). In the next three subsections we describe in detail how
three different components account for the observed SED of HD 34282.

\subsubsection{The long wavelength part of the SED}
\label{LONGWAVE}

The flux at 1.3 mm is proportional to the mass of the disk since its
thermal emission is nearly optically thin at this wavelength (see
e.g., Mannings 1994, Natta et al. 1997). Therefore we can use the flux
density at 1.3 mm (Pi\'etu et al. 2003) to estimate the viscosity
parameter $\alpha$ of the system. Under the assumption of steady
accretion, the disk surface density scales as $\Sigma \propto
\dot{M}\alpha^{-1}T^{-1}$, $\Sigma$ being the surface density,
$\dot{M}$ the mass accretion rate, $\alpha$ the viscosity parameter,
and $T$ the temperature of the mid-plane disk (D'Alessio et
al. 1999). For moderate accretion rates, the heating of the disk is
dominated by stellar irradiation and $T$ will not change very much
with the mass accretion rate, so the mass of the disk will be roughly
constant for constant $\dot{M}\alpha^{-1}$. Usual values of $\alpha$
range from 0.1 to 0.001, 0.01 being the one that reproduces best the
millimetre fluxes in T Tauri stars (D'Alessio et al. 2001). We find
that $\alpha\!=\!0.0008$ yields the appropriate millimetre flux for HD
34282, given the stellar and disk parameters considered. 

The slope of the SED at millimetre wavelengths $\alpha_{\rm mm}$ is
normally modelled with the $\beta$ dependence of the opacity
$\kappa_\lambda=\kappa_0 (\lambda/\lambda_0)^{-\beta}$ (Beckwith et
al. 1990; Beckwith \& Sargent 1991) where $\alpha_{\rm mm}=\beta+2$ to
account for the growth of the dust grains. In our case, the millimetre
slope is controlled by the dust distribution and the optical depth of
the disk. To obtain the observed millimetre slope $\alpha_{\rm
mm}\!=\!2.2$ ($F_\nu\!\propto\!\nu^{2.2}$), a combination of
$p\!=\!2.5$ and a maximum grain size $a_{\rm max}\!=\!1$ cm was
needed. This value of $p$ can be interpreted as the resulting size
distribution of a coagulation process in which most of the disk dust
mass resides in large grains (Miyake \& Nakagawa 1993).

The contribution of the {\it mid-plane disk} to the SED is shown in
Fig. \ref{hd34282fit} as a long-dashed line. There is an excess flux
with respect to the observed flux at 3.4 mm measured by Pi\'etu et
al. (2003). We consider the sub-millimetre and millimetre slope well
fitted with $F_\nu \propto \nu^{2.2}$ since five flux densities,
between 0.45 and 2.6 mm, are consistent with it, and no significant
change in the slope is expected (see, D'Alessio et al. 1990, but also
Qi et al. 2003 for a counter-example). Missing flux, a sudden change
in the dust opacity at large wavelengths or the contribution of
optically thick regions at small radii may explain the discrepancy
(e.g. D'Alessio et al. 2003). A way to check the validity of this
result would be to simulate the observation of this model for the
mid-plane disk in the same line observed by Pi\'etu et al. (2003),
namely the $^{12}$CO $J\!=\!2\rightarrow 1$ transition, and compare
the resulting map with the observations. This work is being done and
it will be presented elsewhere.

\subsubsection{The mid-infrared region of the SED}
\label{MIDIR}

\begin{figure}[h]
\epsfig{file=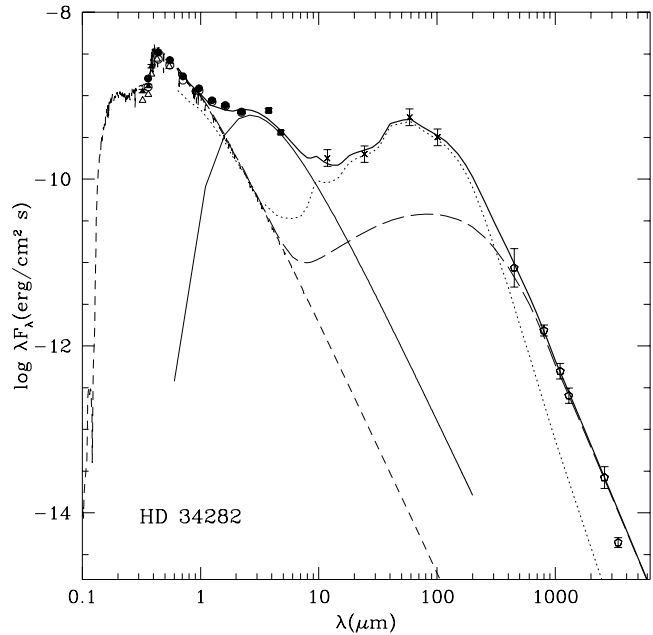, height=9.0cm}
\caption[]{Disk models fitting the full SED of HD 34282. Filled
  triangles are dereddened Walraven photometry from de Geus et al.  (1990).
  Filled circles are dereddened EXPORT optical and near-IR photometry. Filled
  squares are {\it LM} magnitudes from Sylvester et al. (1996). Open
  symbols are the observed fluxes. The crosses are the IRAS
  colour-corrected fluxes and the open pentagons are submillimetre fluxes
  from Sylvester et al.  (1996), Pi\'etu et al. (2003) and Mannings \&
  Sargent (2000).  The dashed line is the Kurucz model for the stellar
  parameters computed in Section \ref{THEPARAMS}. The thin solid line is
  the 1400 K black body fit to the near-IR excess. The dotted line is the
  contribution of the surface model with inclination angle $i=56^{\circ}$,
  $\dot{M}=8.2\times10^{-9}$ M$_\odot$yr$^{-1}$, $\alpha=0.008$, $R_{\rm
    in}$=0.31 AU, $R_{\rm out}$=705 AU, $p=3.5$ and $a_{\rm max}=1$ $\mu$m.
  The long-dashed line is the contribution of the mid-plane disk model with
  the same parameters as the surface one except $p=2.5$, $a_{\rm max}=1$ cm
  and $\alpha=0.0008$. The thick solid line on the top of the others is the
  addition of all components.  See the text for explanation.}
\label{hd34282fit}
\end{figure}

As can be seen in Fig. \ref{hd34282fit} the IR emission coming from
a single irradiated accretion disk model does not reproduce the
observed mid-IR fluxes. According to the CIRR values in the {\it IRAS}
PSC, the cirrus infrared emission around this star is not an important
contribution to the measured fluxes at 60 and 100 $\mu$m, so we
consider that the flux in this part of the SED is also coming from the
disk, implying that a second component is required.

The IR excess produced by an irradiated accretion disk model with dust
and gas well mixed decreases as the size of the dust particles
increases (D'Alessio et al. 2001). For a fixed dust mass, a decrease
in the amount of large dust particles obviously implies the presence
of more small grains in the disk; the effect is that the opacity of
the dust to the stellar radiation increases and the height where most
of the stellar radiation is deposited is greater. In that case, the
disk intercepts more radiation from the star and hence it will emit
more IR flux.  In order to fit the mid-IR excess in HD 34282, we need
the contribution of a disk component that we will call the {\it surface
disk} with $a_{\rm max}\!=\!1\,\mu$m, $p\!=\!3.5$, the same mass
accretion rate as the mid-plane disk (in this way we assure that the
mass transfer is conserved through the disk) and $\alpha=0.008$. The
contribution of the surface disk to the SED is shown in Fig.
\ref{hd34282fit} as a dotted line.

Considering both components, the disk would contain large grains in
the mid-plane, which are responsible for the almost optically thin
emission at sub-millimetre and millimetre wavelengths and a population
of small grains -- the surface disk -- on top of the previous model,
which is responsible for the emission in the mid-IR.  The total mass
of the disk around HD 34282 can now be estimated from the masses of
both disk components (Table \ref{DISKS}) assuming that the mid-plane
disk extends out to some intermediate height and the surface disk
occupies the higher layer to the end of the disk.  Given that the
disks are in hydrostatic equilibrium, the density distribution with
height is a gaussian function centered at the mid-plane. The total
mass is then dominated by the mid-plane disk and lies between 0.6 and
0.7 M$_\odot$. The disk mass is less than half the stellar mass,
supporting the hypothesis that the main gravitational field is due to
the central star.

Natta et al. (2000) used the flux at 2.6 millimetres to estimate a
mass of 0.05 M$_\odot$ for the disk of HD 34282, an order of magnitude
smaller than our value. Differences in the gas temperature and the
opacity parametrization used in their calculation may account
for the difference. Nevertheless, the upper limit of the disk masses
computed by Natta and collaborators for a large sample of HAeBe stars
matches our estimate for the mass of the disk around HD 34282, showing
that it is quite massive compared to typical HAeBe disks.

\subsubsection{The near-IR wavelength region of the disk}
\label{NEARIR}

Some HAeBe stars show conspicuous near-IR bumps. Hillenbrand et al.
(1992) used high mass accretion rate models to fit their SEDs but
Hartmann et al. (1993) argued that such large values for $\dot{M}$
were incompatible with the bolometric luminosity of the objects.
Finally Natta et al. (2001) proposed the presence of a frontally
illuminated wall at the dust destruction radius where this near-IR
bump would be emitted making compatible the shape of the SED and the
luminosity of the object.

We assume that the excess emission at 3 $\mu$m in the SED of HD 34282
originates in a {\it wall-like} structure at the inner part of the disk and
model it as a black body at 1400 K. Following Natta et al. (2001)
and Dullemond et al. (2001), if this wall of irradiated material
were optically thick then it would radiate as a black body at the local
temperature, namely that of dust destruction (according to Pollack et
al. 1994, $T_{\rm evap}\!\sim\!1200 - 1400$ K).

The position, $R_{\rm rim}$, and height, $H_{\rm rim}$, of this irradiated
wall can be computed following the methods described in Dullemond et al.
(2001).  Also, Muzerolle et al. (2003b) have recently fitted the near-IR
bump in the SEDs of Classical T Tauri stars with a black body arising from
a similar model for a dust destruction wall. These authors find best-fit
temperatures for the wall to be around 1400 K. The main difference with
respect to the first model is that the dust is not treated as optically
thick but the position of the rim is computed with the same dust size
distributions as in D'Alessio et al. (2001) and the methods used in Calvet
et al. (1991, 1992).

The projected area, $A$, of the rim is computed as in appendix B in
Dullemond et al. (2001). The solid angle subtended by this area is
$A/d^{2}$, where $d$ is the distance to the source. The normalization
constant needed to fit a black body to the near-IR photometry is
proportional to the solid angle subtended by the rim, so after 
computing the position of the wall, $R_{\rm rim}$, we can estimate
the corresponding height so that the emitting area is consistent
with the photometry.

Then, using equation (14) in Dullemond et al. (2001) in a simple
iterative procedure, we get $R_{\rm rim}\!=\!0.31$ AU and $H_{\rm
rim}\!=\!0.06$ AU.  Also, using the equation for $R_{\rm rim}$ in
Muzerolle et al. (2003b) similar results were obtained (namely $R_{\rm
rim}\!=\!0.31$ AU and $H_{\rm rim}\!=\!0.04$ AU). This agreement is
due to the fact that the different optical depth assumed for the dust
in Muzerolle et al. (2003) is compensated by the autoirradiation
correction introduced by Dullemond et al. (2001).

It is remarkable that the height obtained in this way is similar to
the height of the surface disk component at $R_{\rm rim}$ ($H_{\rm
model}\!=\!0.06$ AU). However, we must take into account that the
height of a single disk where the large grains are in the mid-plane
and the small grains in upper layers may be different to
this\footnote{This is a consequence of the fact that the model with
large $a_{\rm max}$ has a smaller vertical scale height -- is less
flared -- than the model with small $a_{\rm max}$ (see Table
\ref{DISKS}), so the dust in the composite model would look like a
slighly flared distribution of large particles surrounded by a more
flared structure of smaller particles.}. Firmer conclusions about
these heights require a self-consistent determination of the vertical
structure with settled dust, which is left for future work.

As quoted in Natta et al. (2001), the hypothesis of the irradiated
wall assumes a negligible opacity of the gas disk inside $R_{\rm rim}$
so that the stellar radiation is able to reach the inner wall. The
calculation of the actual gas opacity inside the wall is difficult
because of the multiple atomic and molecular absorption lines
present. Muzerolle et al. (2003b) studied the opacity of a gaseous
disk inside the dust destruction radius using an extended set of gas
opacities and a detailed treatment of the heating of the gas due to
accretion and stellar irradiation. They found that for typical
parameters of HAeBe stars, the optical depth of the inner hole would
be low enough to allow a substantial fraction of the stellar radiation
to reach the dust sublimation wall. 

The thin solid line in Fig. \ref{hd34282fit} is the contribution of
the irradiated wall to the SED. The thick solid line is the addition
of the Kurucz model and the contributions to the IR excess of the wall
and the mid-plane and surface disks. In Table \ref{DISKS} a summary of
the input parameters for the models is given. Also the values found
for the height at 10 AU and the disks' masses for each component are
given.

\begin{table}[h]
  \caption[]{Disk model parameters and results.}
   \begin{tabular}{lccc}
                                       &\multicolumn{2}{c}{HD 34282}                     & {HD 141569}             \\\hline
                                       & {\it Mid-plane}      &  {\it Surface}           &                         \\\hline
\rb{$\dot{M}$ (M$_\odot$ yr$^{-1}$)}   & \rb{$8\times10^{-9}$}&  \rb{$8\times10^{-9}$}   & \rb{$1\times10^{-11}$}  \\
$\alpha$                               & $8\times10^{-4}$     &  $8\times10^{-3}$        & $1\times10^{-3}$        \\
$R_{\rm in}$ (AU)                      & 0.31                 & 0.31                     & 0.24                    \\
$R_{\rm out}$ (AU)                     & 705                  & 705                      & 428                     \\
$p$                                    & 2.5                  & 3.5                      & 2.5                     \\
$a_{\rm max}$                          & 1 cm                 & 1 $\mu$m                 & 1 mm                    \\
Inclination                            & 56$^\circ$           & 56$^\circ$               & 51$^\circ$              \\\hline  
$H_{\rm 10\;AU}$ (AU)                  & 3.06                 & 4.13                     & 2.33                    \\
$M_{\rm disk}$ (M$_\odot$)             & 0.70                & 0.06                    & $6.3\times10^{-4}$      \\
\hline
   \end{tabular}
\label{DISKS}
\end{table}

\subsection{The disk model for HD 141569}
\label{MODELHD141569}

The SED of this star has already been fitted with optically thin dust
models (e.g.  Sylvester et al. 1996, Malfait et al. 1998).  Recently Li \&
Lunine (2003) have published a highly detailed, thorough model in which the
star is taken to be a Vega-type object with a debris disk filled with
optically thin porous dust. We have shown in previous sections, however,
that HD 141569 is a PMS HAeBe star and not a Vega-type MS star, surrounded
by a CS disk. This disk is clearly more evolved than usual CS disks
detected around other HAeBe stars, but it cannot be considered as a
`typical' debris disk, as indicated by the ratio between the IR excess and
the bolometric luminosity and the non-negligible gas content and activity.
Thus, the HD 141569 disk appears to be in an intermediate stage between
`first generation' PMS disks and `second generation' debris disks.  We take
this as a justification to carry out an exploratory study in which we fit
the SED of HD 141569 with PMS disk models and then compare the results with
those from the debris disks models, looking for convergent aspects. Note
also that A-type Vega-like stars with debris disks are older systems than
HD 141569 (Song et al. 2001).

The observational results strongly indicate that there is no significant
mass accretion. Thus, we assume that the disk is heated only by irradiation
from the central star and that no viscous dissipation from the mass
accretion onto the star is present, which translates in the models we are
using into an extremely low mass accretion rate. The central star has the
following parameters (Section \ref{THEPARAMS}): $T_{\rm eff}$ = 10000 K,
$R_*\!=\!1.70$ R$_\odot$, $M_*\!=\!2.00$ M$_\odot$, $L_*\!=\!25.77$
L$_\odot$ and $d\!=\!108$ pc, where the stellar radius was calculated from
the stellar mass and the surface gravity. For the disk, we will use an
inclination angle $i\!=\!51^\circ$ (Weinberger et al. 1999) and an outer
radius $R_{\rm out}\!=\!428$ AU, i.e. the value of 400 AU given by
Weinberger et al.  (1999) scaled to our estimated distance of 108 pc. A
change in the outer radius of no more than 50 - 60 AU will not affect
substantially the final results. The inner radius is fitted to be $R_{\rm
  in}$=0.24 AU to account for the lack of near-IR excess in the SED.

\begin{figure}
\epsfig{file=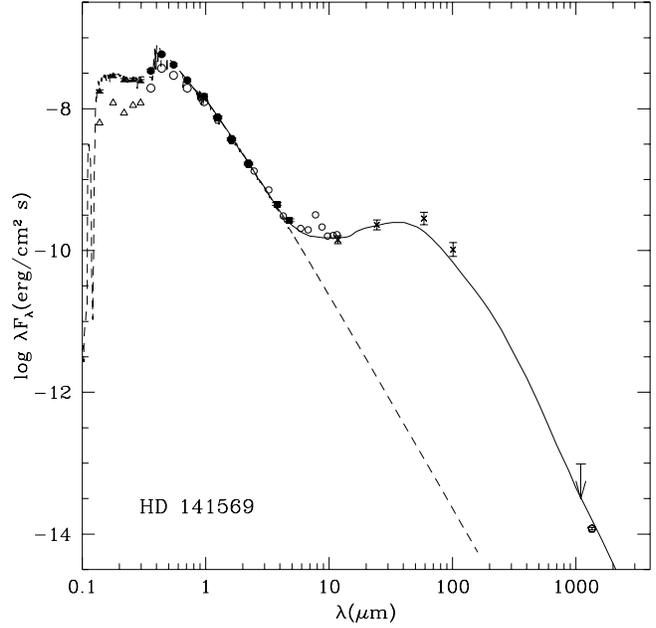, height=9.0cm}
\caption[]{Disk model fitting the SED of HD 141569. Filled triangles
  are dereddened {\it IUE} data. Filled circles are dereddened EXPORT optical
  magnitudes. Filled hexagons are dereddened {\it JHK} magnitudes from
  Malfait et al. (1998) and filled squares are {\it LM} magnitudes from
  Sylvester et al. (1996).  The corresponding open symbols are the observed
  fluxes. The small open circles are the {\it ISO}-PHOT measurements. The
  crosses are the {\it IRAS} colour-corrected fluxes, the open pentagon is
  the submillimetre flux from Sylvester et al. (2001) and the arrow is the
  upper limit flux from Sylvester et al. (1996). The dashed line is the
  Kurucz model for the stellar parameters computed in Section
  \ref{THEPARAMS}. The thick solid line is the disk model with inclination
  angle $i\!=\!51^{\circ}$, $\dot{M}\!=\!10^{-11}$ M$_\odot$yr$^{-1}$,
  $\alpha\!=\!0.001$, $R_{\rm in}$=0.24 AU, $R_{\rm out}$=428 AU,
  $p\!=\!2.5$ and $a_{\rm max}\!=\!1$ mm. See text for details.}
\label{hd141569fit}
\end{figure}  

Fig. \ref{hd141569fit} shows the results of our irradiated disk model
compared to the SED of HD 141569. The peak at 7.7 $\mu$m is a PAH
feature (see Li \& Lunine 2003) that has not been modelled because it
is not included in the dust model used (D'Alessio et al. 2001, Pollack
et al.  1994). We find that a disk with a negligible mass accretion
rate, $\dot{M}\! \sim\!  10^{-11}$ M$_\odot$ yr$^{-1}$, and
$\alpha=0.001$ (which translates into a dust and gas distribution of
$\Sigma \propto R^{-1.0}$ with $\Sigma(10\,{\rm AU})=3$ g cm$^{-2}$),
and a dust distribution with $a_{\rm max}=1$ mm and $p=2.5$ fits
reasonably well the multiwavelength SED of HD 141569. The fractional
IR luminosity from our model is 0.011, consistent with the measured
value (Section \ref{THEPARAMS}).  The total mass (gas and dust) of our
disk is 6.4 $\times 10^{-4}$ M$_\odot$ which is equivalent to 213
M$_\oplus$. The mass of the disk derived with this model is higher
than the typical masses encountered in Vega-like disks (Thi et al.
2001) but clearly smaller than those found in HAeBe stars (Natta et
al. 1997), as expected given the evolutionary status of HD 141569.
Table \ref{DISKS} gives a summary of the input parameters for the
model as well as the mass of the disk and the height at 10
AU\footnote{Note that, whereas the disk of HD 141569 seems to be more
  evolved than that of HD 34282, the estimated age of HD 141569 is
  {\it less} than that of HD 34282 (see Table \ref{STARS}). However,
  one has to be careful when comparing this parameter for both stars,
  given the fact that HD 141569 is more massive and has larger metal
  abundance than HD 34282.  Therefore, the evolutionary time scales
  are different for both objects. The initial conditions of the clouds
  where the stars and disks formed must also play a role in the
  evolutionary scenario.}.

In the model by Li \& Lunine (2003) the surface density of dust
particles is introduced with three analytical approximations that
reproduce the surface brightness profiles observed in thermal and
scattered light images by Augereau et al. (1999), Weinberger et al.
(1999), Mouillet et al. (2001) and Marsh et al. (2002). However, the
vertical height of the dust is computed assuming vertical hydrostatic
equilibrium with a gas temperature taken to be a power law of the
distance from the central star. In our disk models, both the density
profile and the vertical scale height are computed taking into account
the stellar irradiation heating (assuming that dust and gas are in
thermal equilibrium). Concerning the continuum emission in the
infrared to millimetre wavelengths, the presence of gaps of moderate
width with smaller dust density does not change dramatically the shape
of the SED (see e.g. Fig.  3 in Li \& Lunine (2003) where the three
dust components are added to yield the smooth observed SED). Therefore
we can assume that our continuous dust distribution is a good first
approximation to the actual dust density with the advantage that it is
self-consistently computed, along with its vertical hydrostatic
structure, with the set of stellar parameters previously determined.
This is in agreement with the results obtained by Clampin et al.
(2003).

Concerning the dust, Li \& Lunine (2003) found that a
`cold-coagulation' dust composed of porous dust grains with a porosity
$P\!=\!0.9$ (90 \% of the volume is vacuum and 10 \% material), and a
power law distribution of sizes $n(a) \propto a^{-3.3}$ with 1 $\mu$m
$< a <$ 1 cm, fitted very accurately the whole continuum SED of HD
141569. This dust model is not very far from our assumed $n(a) \propto
a^{-2.5}$ with 0.005 $\mu$m $< a <$ 1 mm mostly at large sizes. The
difference in the optical properties of the compact dust grains
compared to the porous dust grains may be accounted for by using a
different power law exponent for the size distribution. The
contribution of the PAH molecules to the IR emission is more important
around the peaks at 3.3, 6.6, 7.7, 11.3 $\mu$m (see Li \& Draine 2001
for details) and the continuum is well reproduced by the porous or
compact dust populations.

The main effect of having a disk with gas and dust, instead of a pure
dusty disk, is that the heating due to stellar irradiation produces an
increase in the vertical scale height of the disk, yielding a more
dilute dust population. Summarizing, we find that our approach fits
the observed SED of HD 141569 reasonably well, although the model we
have used does not include as many details as that developed by Li \&
Lunine, particulary in the treatment of the dust properties. In this
sense, our result is interesting since it shows the possibility of
studying CS disks in transition objects from both perspectives, namely
the PMS and MS sides.

\section{Conclusions}
\label{THECONC}

We have presented a study of the properties and spectral energy
distributions of the HAeBe stars HD 34282 and HD 141569 and their disks,
based mainly on observations made by the EXPORT consortium (Eiroa et al. 2000).
The main conclusions can be summarized as follows: 

\begin{itemize}
  
\item[--]The spectroscopic and photometric observations of both stars
  have been described and analysed in detail with the aim of throwing
  light on their evolutionary status and the dynamics and variability
  of their photospheres and disks.
  
\item[--]Both stars are metal-deficient. Our estimates are [Fe/H]=$-0.8$ for
 HD 34282 and $-0.5$ for HD 141569. 
  
\item[--]The analysis has provided values for the stellar distances.
  The distance to HD 34282, namely 348 pc, is consistent with the
  value given by Pi\'etu et al. (2003) and corrects the {\it Hipparcos}
  distance that is uncertain due to the large error in the parallax.
  The new distance for HD 34282, together with its low metallicity,
  resolves the anomalous position of this star in the HR diagram
  previously reported under the assumption that the star has solar
  abundances. The distance to HD 141569, namely 108 pc, is consistent
  with the {\it Hipparcos} data and in this case, the question of the
  anomalous position of this star in the HR diagram is resolved by
  using the correct set of tracks with the appropriate metallicity.
  
\item[--]In addition to the above two parameters, our analysis has provided
  for both stars values for the effective temperatures, spectral types,
  masses, gravities, luminosities, ages, projected rotational velocities
  and extinctions.  These results are given in Table \ref{STARS}.
  
\item[--]Complete SEDs from the ultraviolet to the millimetre range have
  been constructed for both stars and their disks using the EXPORT data plus
  additional results gathered from the literature and on-line databases.
  Table \ref{TableSED} gives a full account of this compilation and the
  data sources.
  
\item[--]The self-consistent irradiated accretion disk models of
  D'Alessio et al. (1998, 1999 and 2001) have been used to fit the SEDs of
  both systems. A summary of the disk parameters obtained is given in 
  Table \ref{DISKS}.
  
\item[--]The SED of the disk of HD 34282 has been reproduced using a
  three-component model. The {\it mid-plane disk} is modelled with a maximum
  dust grain size of 1 cm and emits mainly at submillimetre and millimetre
  wavelengths, the {\it surface disk} model has a maximum grain size of 1
  $\mu$m and is responsible for the mid-IR excess emission. A near-IR bump
  similar to those seen in other HAeBe stars (see Natta et al. 2001) is
  fitted with the emission of a frontally illuminated {\it wall} placed at the
  dust destruction radius (Dullemond et al.  2001; Muzerolle et al. 2003b, 
  D'Alessio et al. 2003). The geometry of the wall is constrained by the
  near-IR photometry.
  
\item[--]In the process of fitting the SED of HD 34282, we have tried
  for the first time a new method to estimate the mass accretion rate
  towards a HAeBe star using ultraviolet Walraven photometry and the
  magnetospheric accretion shock models of Muzerolle et al. (2003a).
  We obtained un upper limit for the mass accretion rate in HD 34282
  compatible with the parameters needed to fit the millimetre SED.
  The low value found does not support the idea by Herbst \&
  Shevchenko (1999) about the FU Ori-like mechanism causing the
  photometric activity of UX Ori stars.
  
\item[--]The SED of HD 141569 has been fitted with a disk model
  irradiated by the central star with negligible accretion and with
  large grains ($a_{\rm max}=1$ mm). A slightly flared dust and gas
  distribution was obtained, which is consistent with a substantial
  amount of residual gas in the disk.

\end{itemize}

\begin{acknowledgements}
  We would like to acknowledge Prof. Vladimir Grinin for his careful
  reading of the manuscript and his constructive suggestions.
  The authors are also grateful to Sukyoung Yi for kindly providing
  unpublished smoothed evolutionary tracks and for advice on their
  use, and to David Barrado y Navascu\'es, Pedro Garc\'{\i}a-Lario,
  Jos\'e Francisco G\'omez and Pedro Rodr\'{\i}guez for discussions.
  The work of C. Eiroa, B. Mer\'{\i}n, B. Montesinos, A. Mora and E.
  Solano has been supported in part by the Spanish grant
  AYA2001-1124-C02. P. D'Alessio acknowledges grants from DGAPA-UNAM and
  CONACyT, M\'exico. B. Mer\'{\i}n wishes to acknowledge the
  hospitality of the Young Stars' group at the Harvard-Smithsonian
  Center for Astrophysics and its support with the STScI grant GO-9524
  during his stay as a visiting student, and to the INTA for its
  financial support of a graduate fellowship.
\end{acknowledgements}

\end{document}